\documentclass[aps,prb,reprint]{revtex4-1}
\usepackage{epsf}
\usepackage{amsfonts}
\usepackage[fleqn]{amsmath}
\usepackage{amssymb,yfonts,mathrsfs,bbm}
\usepackage{graphicx}
\usepackage{multirow}
\usepackage{longtable}
\usepackage{amsmath}
\usepackage[normal]{caption2}

\usepackage{color}
\definecolor{purple}{rgb}{0.7,0.0,0.7}
\definecolor{orange}{rgb}{1,0.65,0.0}
\definecolor{dgreen}{rgb}{0.3, 0.6, 0.1}

\begin{document}
\title{Observation of forbidden phonons and dark excitons by resonance Raman scattering in few-layer WS$_2$}

\author{Qing-Hai Tan$^{1,2}$}
\author{Yu-Jia Sun$^{1,2}$}
\author{Xue-Lu Liu$^{1,2}$}
\author{Yanyuan Zhao$^{3}$}
\author{Qihua Xiong$^{3}$}
\author{Ping-Heng Tan$^{1,2*}$}
\author{Jun Zhang$^{1,2*}$}
\affiliation{$^{1}$State Key Laboratory of Superlattices and Microstructures, Institute of Semiconductors, Chinese Academy of Sciences, Beijing 100083, China
\\$^{2}$College of Materials Science and Opto-Electronic Technology, University of Chinese Academy of Sciences, Beijing 101408, China
\\$^{3}$Division of Physics and Applied Physics, School of Physical and Mathematical Sciences, Nanyang Technological University, Singapore 637371, Singapore
\\*Correspondence and requests for materials should be addressed to P. T. and J. Z. (Email:  phtan@semi.ac.cn, zhangjwill@semi.ac.cn)}

\begin{abstract}
{The optical properties of the two-dimensional (2D) crystals are dominated by tightly bound electron-hole pairs (excitons) and lattice vibration modes (phonons). The exciton-phonon interaction is fundamentally important to understand the optical properties of 2D materials and thus help develop emerging 2D crystal based optoelectronic devices. Here, we presented the excitonic resonant Raman scattering (RRS) spectra of few-layer WS$_2$ excited by 11 lasers lines covered all of A, B and C exciton transition energies at different sample temperatures from 4 to 300 K. As a result, we are not only able to probe the forbidden phonon modes unobserved in ordinary Raman scattering, but also can determine the bright and dark state fine structures of 1s A exciton. In particular, we also observed the quantum interference between low-energy discrete phonon and exciton continuum under resonant excitation. Our works pave a way to understand the exciton-phonon coupling and many-body effects in 2D materials.}\end{abstract}

\maketitle
The exciton-phonon interaction plays an important role in determination of optical properties in semiconductors and related optoelectronic devices performances. For instance, strong exciton-phonon coupling can strongly enhance the phonon-assisted anti-Stokes luminescence upconversion and leads to net optical refrigeration of semiconductors\cite{ZJ-Nature-2013}. Particularly, when the exciton-phonon coupling is only efficient between specific phonon branch and exciton state, many interesting optical phenomena are emerged such as resolved sideband Raman cooling of optical phonons in semiconductors\cite{ZJ-NP-2016}, phonon-exciton polariton\cite{polaritons-prb-1979}, and multi-phonon replica of exciton hot-luminescence\cite{CdS-PRL}. In those experiments, resonant Raman scattering (RRS) was extensively used to detect and study the exciton-phonon interaction.

Two dimensional (2D) Layered transition metal dichalcogenides (TMDs) with stoichiometry of MX$_2$ (M=Mo, W; X=S, Se, Te) offer a platform with both strong exciton-photon and exciton-phonon interaction effects\cite{PhysRevB-85115317-2012,Carvalho-prl-2015}. In few-layer MX$_2$, owing to the dielectric screening and spatial quantum confinement in 2D systems, their optical transitions and light-matter interaction are governed by robust exciton feature with binding energies of several hundred meV \cite{um-nm-2014,Qiu-PRL-2013,cuixiaodong-serp-2015}. Due to strong spin-orbit (SO) coupling and exciton-phonon interaction, it allows to study the spin-valley circular dichroism\cite{Xu-NPRiew-2016,Cao-NC-2012,cuixiaodong-PNAS-2014,cuixiaodong-review-2015}, dark exciton\cite{Dark-exciton-nature-2014}, phonon-assisted excitonic luminescence upconversion \cite{NP-xuxiaodong-2016}, valley selective exciton-phonon scattering\cite{mm-arXiv-2017}, and dipole-type polaritonic excitations \cite{NM-LowT-2017}.

Raman scattering is a powerful tool to detect the phonon dynamics 2D materials\cite{Zhang-Xin-CSR-2015,Lux-nanoresarch-2016}. In particular, Raman spectroscopy can be used to detect the interlayer coupling and exactly determine the layer number of 2D sample \cite{Tan-nm-2012,Zhao-nanoletter-2013}. When the laser energy is close to the exciton transitions of TMDs, RRS allows people to explore the exciton-phonon coupling\cite{Sun-PRL-2013,del-corro-nanolett-2014,MAA-PRB-2014,mote2-prb-2016,MoSe2-prb-2016,Wu-nc-2014}, symmetry-dependent properties of exciton-phonon scattering \cite{Carvalho-prl-2015, del-nanolett-2016}, phonon assisted intervalley scattering \cite{CBR-Naturecom-2017,PhysRevLett-117-2016, WSe2-prl-2015}, $\emph{etc}$. Up to now, most of RRS experiments in MX$_2$ only focus on high frequencies Raman modes of MX$_2$, while the ultra-low frequency phonon RRS (below 50 cm$^{-1}$) has rarely been reported. Since ultra-low frequency phonon modes have a very small energy of several meV, we can use it to explore the fine-structure of excitonic energy sates such as small splitting excitonic levels. Secondly, some forbidden phonon modes and excitonic states become observable and active in RRS experiments based on parity selection rule\cite{cu2o-prb-1973,cu2o-prl-1973}: in ordinary non resonance Raman scattering, only Raman active phonons with even parity can be observed but infrared (IR) active phonons with odd parity cannot be observed; however, in RRS situation, the odd-parity IR-active phonons can be activated when the Raman transition includes the electronic-dipole forbidden excitons with even parity (dark excitons). Since the dark and bright excitons governs the optical properties of 2D TMDs\cite{Dark-exciton-nature-2014} and affects the performances of 2D based opto-electronic devices\cite{Yeyu-Natphoton-2015}, it is thus crucial to detect the bright-dark exciton fine-structures and determine their splitting energy in TMDs. Recently there are several experimental techniques such as two-photon absorption/photoluminescence and temperature dependent time-resolved photoluminescence have been used to probe the dark excitons in TMDs \cite{Dark-exciton-nature-2014, trion-NC-2016, ZhangXX-prl-2015}. However, those methods can't detect the forbidden-phonon and dark excitons simultaneously and thus can't provide detail information of the exciton-phonon coupling related to forbidden phonons and dark excitons.

Furthermore, when the discrete phonon mode couples with a continuum state such as electron occupation states, the phonon lineshape became an asymmetry Breit-Wigner-Fano (BWF) resonance lineshape (in short Fano resonance), which is fundamental important in the interpretation of light propagation, optical spectra and electronic transport in semiconductors\cite{Fano-pr-1961,RevModPhys-fano-review,FANORESONANCE-NM-2010}. The Fano resonance of phonon Raman spectra has been widely studied in doped silicon\cite{semifano-prb-1975}, graphitic materials\cite{nanotube-nature-1997,PhysRevLett-graphene-2009,Tan-nm-2012}, and layered topological insulator\cite{ZJ-nanolett-2011}, charge density wave and superconductor states\cite{RevModPhys-fano-review,RevModPhys-79-75}, $\emph{etc}$. However, to the best of our knowledge, there are few reports about the exciton-phonon Fano resonance in MX$_2$.

Here, we investigated forbidden phonons, dark excitons and exciton-phonon interactions in few-layer WS$_2$ by using of RRS techniques up to 11 excitation laser lines to finely resonate with A, B and C exciton transition from 4 to 300 K.

\begin{figure*}[htb]
\centerline{\includegraphics[width=170mm,clip]{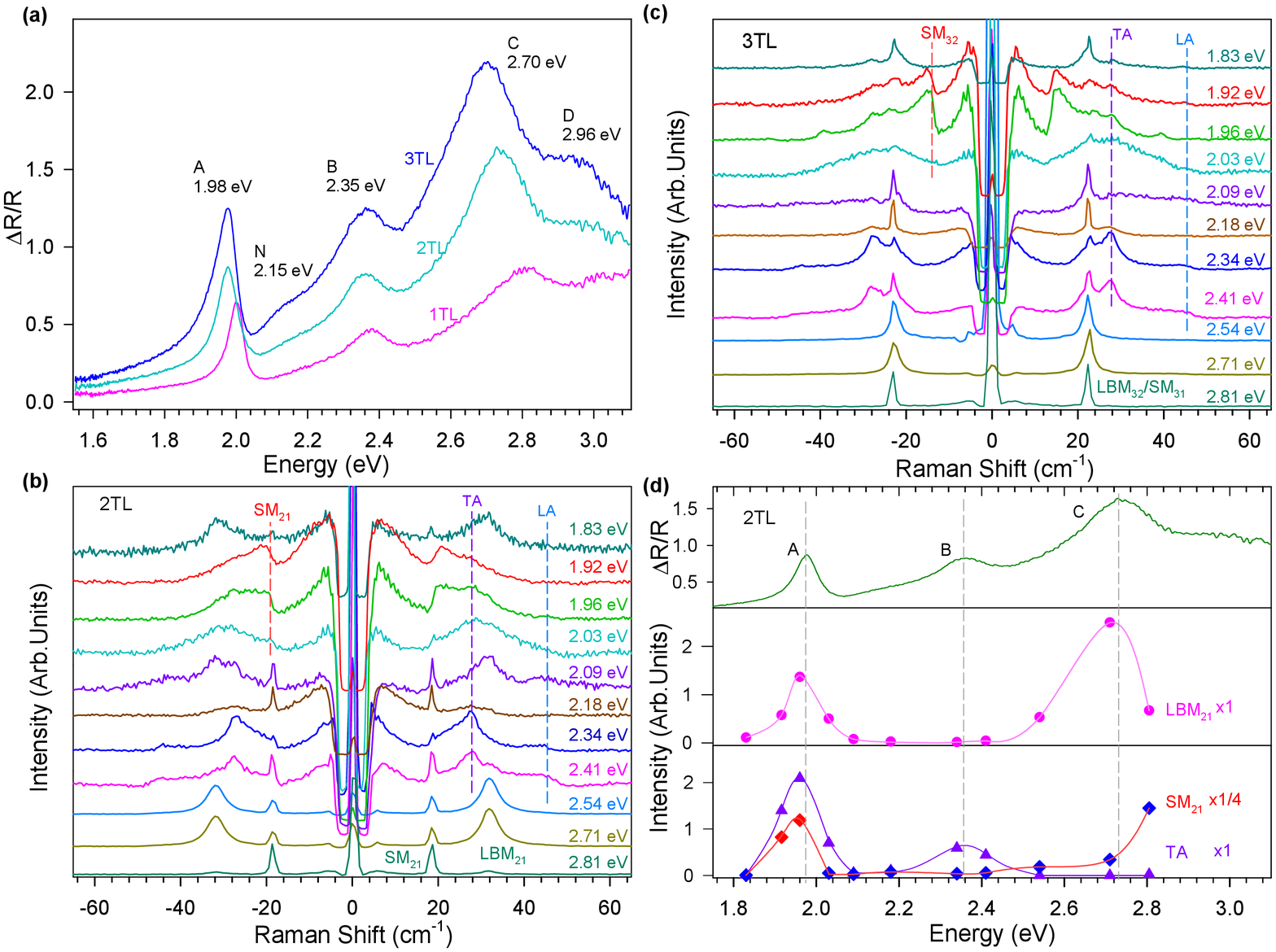}}
\caption{\textbf{Reflectance contrast spectra $\Delta{R/R}$ and ultra-low frequency RRS spectra of few-layer WS$_2$.} (a), Reflectance contrast spectra $\Delta{R/R}$ for 1TL, 2TL, and 3TL WS$_2$ crystals on SiO$_2$ substrate. (b-c), Ultra-low frequency Raman spectra of 2-3TL WS$_2$ samples with different laser excitation energy (E$_L$). (d), Reflectance contrast spectra $\Delta{R/R}$ for 2TL WS$_2$ (upper) and RRS profiles of the layer breathing mode (LBM$_{21}$) shear modes (SM$_{21}$) (Fano lineshape, red diamonds; Lorentzian lineshape, blue diamonds), and TA modes in (b). The dash lines are used to guide to the eyes.} \label{Fig1}
\end{figure*}
\vspace*{5mm}

\subsection*{Results}

Like other MX$_2$ TMDs, WS$_2$ shares the same hexagonal crystalline structure. Each layer is composed of three atomic layers, one W layer sandwiched between two S layers, and we therefore refer them as a trilayer (TL). The W and S atoms could occupy different trigonal sites, forming different polytypes. For Mo and W dichalcogenides, 2H (here H stands for hexagonal) is the most stable and common polytype in nature, where two adjacent atomic layers keep alternating at A and B sites.

Figure 1(a) shows the reflectance contrast spectra $\Delta{R}/R$ for 1TL, 2TL, and 3TL WS$_2$ on SiO$_2$ substrate at room temperature. Two features around 1.98 and 2.35 eV, denoted by A and B exciton, is attributed to the interband dipole-allowed exciton transitions between the spin-orbit split valence band and the lowest conduction at K (or K') point of the Brillouin zone, respectively\cite{Mark-F-prl-2010,LYL-PRB-2014}. The broad peak around 2.7 eV, denoted by C exciton, is from transition between the highest valence band and the first three lowest conduction bands around the $\Gamma$ point of the Brillouin zone\cite{Qiu-PRL-2013}. While the features N and D peaks around 2.15 and 3.0 eV are not clear so far. Both the energies of A and B exciton in 1TL are slightly larger than that in 2TL and 3TL WS$_2$. In contrast to A and B exciton transitions, the energy of C exciton transitions red shifts to 2.7 eV in 3TL from around 2.8 eV in 1TL.

Since the energy of each exciton in WS$_2$ is distinct from each other, it provides a perfect platform to study the excitonic RRS behavior in WS$_2$. Figure 1(b) and 1(c) shows the ultra-low frequency RRS spectra excited by 11 laser lines on 2TL and 3TL WS$_2$, respectively. When the excitation energies ($\varepsilon_{ex}$) are resonant with C exciton transitions in the range of 2.81 to 2.54 eV, we only observed the interlayer shear modes (SMs) and interlayer breathing modes (LBMs), of which frequencies are strictly depend on the layer numbers\cite{Tan-nm-2012,Zhang-prb-2013,Zhao-nanoletter-2013}. We also measured the Raman spectra of 1-5TL, 7TL, and bulk WS$_2$ excited by 2.71 eV laser as shown in supporting Figure S1(a). There are no new modes below 50 cm$^{-1}$ been found expect SMs and LBMs, of which frequencies match very well with theoretical results calculated by liner chain model (LCM)\cite{lixl-nt-2016}, as shown in supporting Figure S1(b). Here we define the SM$_{N1}$ is the highest frequency branch and SM$_{NN-1}$ is the lowest frequency branch of the SMs, where N refers to the layer numbers of sample and N-1 refers to the (N-1)th branch of SMs\cite{Zhao-nanoletter-2013,Zhang-Xin-CSR-2015}. Similarly, the LBM$_{N1}$ and LBM$_{NN-1}$ are the highest and lowest frequency branch of the LBMs, respectively. Therefore, we can use those modes as a fingerprint to unambiguously identify the layer numbers of WS$_2$.

When the $\varepsilon_{ex}$ is close to B exciton transitions (2.41 to 2.18 eV), besides the SM and LBM, we also observed two new peaks located at 27.8 and 45.4 cm$^{-1}$, which frequencies are independent on the $\varepsilon_{ex}$ and layer numbers. In 2TL and 3TL, the peak at 45.4 cm$^{-1}$ is not so pronounced as 27.8 cm$^{-1}$ peak, but we can still resolve it. As being discussed in later part, we tentatively assign them as acoustic phonons with finite wavevector $\emph{k}$ and labeled them as TA and LA, respectively. While the excitation energy is around 2.09 eV, in the middle of A and B exciton transitions, these two wavelength-independent peaks almost vanished. When the $\varepsilon_{ex}$ is close to A exciton transitions (1.92 and 1.96 eV), we observed another new peak with asymmetric Fano lineshape. We noted that its frequency is close to SM$_{NN-1}$. Therefore, we proposed that this Fano lineshape mode is from the interference between SM$_{NN-1}$ phonons and continuum electronic states when $\varepsilon_{ex}$ is resonant with A exciton transition. We will give more evidences to certify this assignment in later part.

In Figure 1(d), we plotted the RRS profiles of three representative Raman modes of LBM$_{21}$, TA and SM$_{21}$ of 2TL WS$_2$. Interestingly, these three kinds of phonon modes show different resonance profiles. The LBM$_{21}$ shows a resonant enhancement at the A and C exciton transitions; the TA mode is enhanced at both of A and B exciton resonances. More interestingly, the SM$_{21}$ shows a Fano lineshape when incident lasers are close to A exciton transition, while it became a Lorentzian lineshape when incident laser is off-resonant with A exciton. We also measured RRS spectra of other high frequency modes (See Figure S2 in supplementary information). As shown in supporting Figure S2(c), all of phonons contributed from high symmetry points in Brillouin zone, including 2LA(M), $E'(M)-LA(M)$ and $A'_1$(M)-LA(M)\cite{Zhang-Xin-CSR-2015,ShiW-2D-2016}, show incident resonant enhancement at A and B exciton transition. Differently, the A$_{1g}$ mode is incidentally resonant enhanced at A, B and C exciton transitions, while E$_{2g}^{1}$ mode is only strongly enhanced at C exciton resonance. This distinct electron-phonon coupling can be explained by considering the symmetries of the phonons and of the exciton orbitals associated with the A, B, and C excitons \cite{Carvalho-prl-2015}.

\begin{figure*}[htb]
\centerline{\includegraphics[width=180mm,clip]{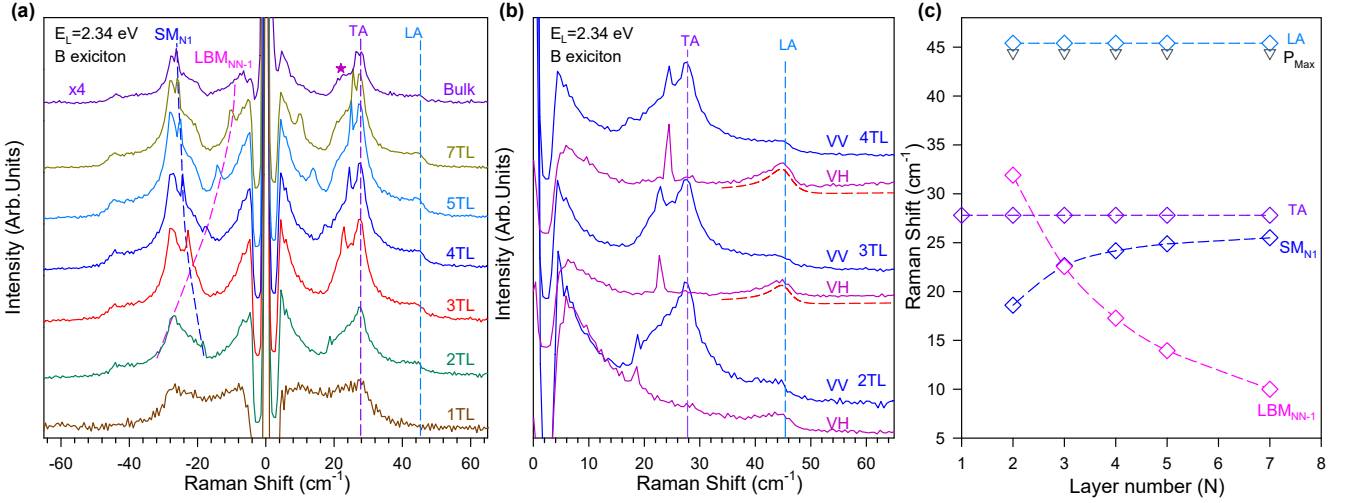}}
\caption{\textbf{RRS spectra of 1-5TL, 7TL and bulk WS$_2$ at B exciton resonance.} (a), RRS spectra of 1-5TL, 7TL and bulk WS$_2$ crystals excited by 2.34 eV laser. The star symbol labeled the unknown peak. (b), Raman spectra of 2-4TL WS$_2$ under parallel (VV) and cross (VH) polarization configurations. The dash curves are the Fano fitting of LA modes, which are offset for clarity. (c), The frequency of SM, LBM, TA, and LA modes depends on the layer numbers of WS$_2$ flakes. Here P$_{Max}$ represent the peak frequency of LA modes. The dash lines are used to guide to the eyes.} \label{Fig2}
\end{figure*}

In order to understand the physical origins and confirm assignments of new observed Raman peaks, we have conducted the RRS measurements at A and B exciton resonance for different layer numbers, polarization Raman configurations, and sample temperature, respectively. Figure 2(a) shows the ultra-low frequency RRS spectra of 1-5TL, 7TL, and bulk WS$_2$ sample excited by 2.34 eV laser, which is close to B exciton transitions. Besides SM$_{N1}$ and LBM$_{NN-1}$ peaks, the 27.8 cm$^{-1}$ mode can be observed for all samples, while the mode at 45.4 cm$^{-1}$ is only absent in monolayer case may due to the too weak scattering cross-section. As shown in Figure 2(b), the 27.8 cm$^{-1}$ mode is polarization dependent, while the mode at 45.4 cm$^{-1}$ is depolarized. Figure 2(c) shows the frequencies of TA, LA, SM$_{N1}$ and LBM$_{NN-1}$ as a function of layer numbers of WS$_2$. Obviously, the frequencies of SM$_{N1}$ and LBM$_{NN-1}$ are strictly depend on the layer numbers, while the frequencies of TA and LA are not only independent on the layer numbers, but the excitation energies, as shown in Figure 2(c) and Figure 1(b-c). We noted that the peak at 45.4 cm$^{-1}$ shows a clearly asymmetrical Fano-like lineshape, which became much more obvious in the case of cross (VH) polarization due to suppressive background signals. The peak at 45.4 cm$^{-1}$ is fitted by Fano function \cite{RevModPhys-fano-review,ZJ-nanolett-2011}:
\begin{equation}
I= I_{0}\frac{(1 + {(\omega-\omega_0)}/{q\gamma})^{2}}{1+(({\omega-\omega_0})/{\gamma})^{2}}
\end{equation}
where the ${\omega_0}$ is uncoupled phonon frequency, $\emph{q}$ is the asymmetry parameter, and $\gamma$ is the full-width-half-maxium (FWHM) of uncoupled phonon modes. In coupled resonance, the peak frequency of Fano lineshape becomes $\omega_0+\gamma/q$ and Fano FWHM becomes 2$\gamma|(q^2+1)/ (q^2-1)|$. For positive $\emph{q}$, the Fano peak frequency is blueshifted related to the uncoupled phonon, but it is redshifted for negative $\emph{q}$. In Fano resonance, the peak linewidth is always broader than intrinsic phonon linewidth no matte positive or negative $\emph{q}$ when $|\emph{q}|>1$. The dimensionless parameter ${1/\emph{q}}$ characterizes electron-phonon coupling strength: a stronger coupling ($\emph{q}$$\to$0) causes the peak to be more asymmetric. In the limit of weak electron-phonon interaction ($\emph{q}$$\to\infty$), Fano lineshape is reduced to a Lorentzian lineshape. Such electron-phonon interaction caused BWF lineshape has been observed in 2D materials, such as graphene\cite{ZhangYB-NN-2010, Tan-nm-2012} and $Bi_2Se_3$ 2D crystals\cite{ZJ-nanolett-2011}. As shown the fitting curve in Figure 2(b), we got $1/q=-0.29$, which is almost a constant for different layer numbers. As shown in Figure 2(c), this negative value of $\emph{q}$ leads to the peak frequencies of LA Fano modes are redshifted related to uncoupled LA phonon, which the frequencies are obtained by Fano fitting. The $|1/\emph{q}|<1$ reflects the 45.4 cm$^{-1}$ phonon mode is dominant in Fano resonance\cite{ZhangYB-NN-2010}.

Based on the lineshape, linewidth and Stokes/anti-Stokes components, we proposed that these modes independent on layer number and excitation energies belong to the lattice phonon rather than other quasi particles such as electronic plasmon. The reason is that the electronic plasmon Raman scattering usually is observed in heave doped sample with very broad single Stokes peak\cite{Abstreiter1984}. Since these two modes are independent on the layer numbers, they may belong to acoustic phonons. As shown in Figure S3 in supplementary information, there are only acoustic phonons in 1TL bellow 60 cm$^{-1}$; in few-layer sample, both of SMs and LBMs are layer number dependent. As results, only acoustic phonons with finite $\emph{k}$ can explain physical origin of these modes. Another evidence for this assignment is to compare the ratio of two mode frequencies and ratio of acoustic velocity related two acoustic phonon branches. We found that the ratio of longitudinal acoustic (LA) velocity (1080 $m/s$) to transvers acoustic (TA) velocity (668 $m/s$) \cite{Phonondispersion-2011} exactly matches the ratio of 45.4 cm$^{-1}$ to 27.8 cm$^{-1}$. Additionally, by comparing the their polarization dependence with Raman tensors of LA and TA vibration\cite{Zhao-nanoletter-2013}, we also found that 45.4 cm$^{-1}$ mode has the same depolarized properties as LA Raman tensor, and TA phonon Raman tensor matches polarized properties of the 27.8 cm$^{-1}$ mode. Therefore, we assigned 27.8 and 45.4 cm$^{-1}$ modes as TA and LA phonons with finite $\emph{k}$ near the Brillouin center, respectively. Based on Figure S3, we can find the $\emph{k}\simeq3\times$10$^{7}$ cm$^{-1}$, which is very close to center of Brillouin zone but much larger than wavevector of incident light ($\emph{k}$$_{light}\simeq10^{5}$ cm$^{-1}$). In resonance condition, the phonons with finite $\emph{k}$ can be involved in the Raman scattering though elastically scattered by a zero energy defect states with -$\emph{k}$. In this case, both of the momentum and energy are conversed. The similar phenomena have been observed in RRS spectra of graphene.\cite{Ferrari-200747} The detailed physical mechanism behind the emerging of these acoustic modes calls for further studies.

\begin{figure*}[htb]
\centerline{\includegraphics[width=180mm,clip]{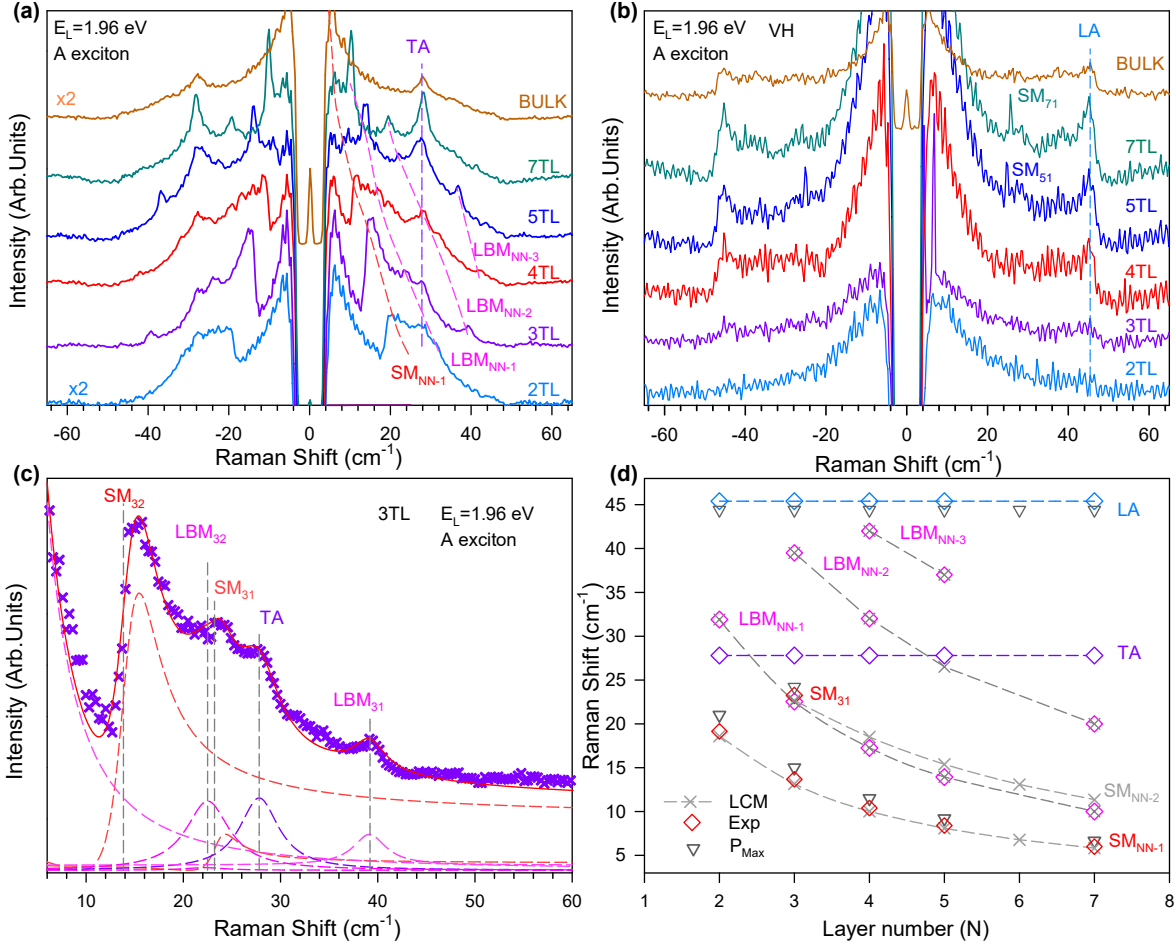}}
\caption{\textbf{RRS spectra of 1-5TL, 7TL and bulk WS$_2$ at A exciton resonance.} (a), Raman spectra of 1-5TL, 7TL and bulk WS$_2$ crystals excited by 1.96 eV laser. (b), Raman spectra of 2-5TL, 7TL and bulk WS$_2$ under cross (VH) polarization configurations. The intensities of Raman modes under VH polarization configurations are around 100 times smaller than that under unpolarized condition in (a). (c), The measured unpolarized Raman spectra and the corresponding fitted results of 3TL WS$_2$. (d), The frequencies of experimental (diamonds) and calculated (gray cross) results as a function of layer numbers of WS$_2$ flakes. LCM represents the liner chain model.} \label{Fig3}
\end{figure*}

We also conducted the RRS experiments resonated with A exciton in 1-5TL, 7TL, and bulk WS$_2$ excited by 1.96 eV and 1.92 eV as shown in Figure 3(a) and supporting Figure S4, respectively. In A exciton resonance case, the TA mode can still be clearly resolved at 27.8 cm$^{-1}$, which has the same frequency as observed at B exciton resonance. However, LA mode is too weak to be resolved, as shown in Figure 3(a). Since the background signal can be efficiently suppressed under VH polarization configuration, we conducted the RRS measurements with the same excitation energy under the VH polarization as shown in Figure 3(b). In contrast to the VH RRS spectra measured at B and C exciton transitions as shown in Figure 2(b), all of SM$_{NN-1}$ modes are disappeared, and SM$_{N1}$ modes can be resolved with a very weak intensity in 5TL and 7TL WS$_2$. The LA mode also shows much weaker intensity compared with spectra under the B exciton resonance condition, but it can still be resolved. The detailed reasons call for further studies. This is another evidence that the exciton-phonon coupling is quite different for A and B excitons.

As shown in Figure 1(a-b), when the incident laser is resonant with A exciton, we found two Fano lineshape peaks close to SM$_{21}$ in 2TL and SM$_{32}$ in 3TL, respectively. We proposed that these Fano peaks are from the quantum interference between SM$_{NN-1}$ modes and electronic continuum state. Exactly, as shown in Figure 3(a), we can still observe these series of layer-dependent Fano resonant SM$_{NN-1}$ modes in all of 2-7TL WS$_2$ samples. The peak intensities in 2-4TL WS$_2$ are much stronger than in thicker samples. Depending on specific resonant exciton properties, different LBMs show different resonant behaviors. Only Raman active LBM$_{NN-1}$ modes appears in B exciton resonance; both Raman active modes of LBM$_{NN-1}$ and LB$_{NN-3}$ (N$>$3) appear in C exciton resonance cases; interestingly, we can observed IR-active LBM$_{NN-2}$ (N$>$2) and Raman-active LBM$_{NN-3}$ (N$>$3) in A exciton resonance. Figure 3(c) shows a fitting result of 3TL WS$_2$ spectrum in Figure 3(a). We should note that the asymmetry direction of Fano resonant peak SM$_{32}$ is towards to higher energy side and gives a positive $\emph{q}$. For the peak around 24 cm$^{-1}$, its frequency is rather close to the SM$_{31}$ mode than LBM$_{32}$, and thus we use Lorentzian LBM$_{32}$ and Fano SM$_{31}$ lineshape to fit the spectra. We concluded all of observed peaks in the Figure 3(d). We found all of LBM frequencies agree the LCM calculation very well. The fitted frequencies of SM$_{31}$ and SM$_{NN-1}$ Fano peaks perfectly match the uncoupled phonon frequencies obtained by LCM, but they are little bit lower than the peak frequencies of corresponding Fano resonant modes. As mentioned previously, this is because the positive $\emph{q}$ of Fano resonance SM$_{NN-1}$ and SM$_{31}$ mode will lead blueshift of the uncoupled SM frequency based on Fano resonance peak function of $\omega_{Fano}=\omega_0+\gamma/q$.  For LA Fano resonant peak in A exciton resonance, we still obtain the negative $\emph{q}$ and redshifted peak frequencies related to uncoupled LA mode.

Based on the group theory\cite{Zhao-nanoletter-2013}, we note that among the N-1 interlayer shear modes of the NTL crystal, only the modes with the j$^{th}$-highest (j=1, 3, 5,...) frequencies (such as SM$_{21}$, SM$_{31}$, SM$_{43}$) can be observed under both the parallel (VV) and VH polarization configurations. The j$^{th}$-highest (j=2, 4, 6, ...) frequency modes have a E$''$ symmetry (Raman-active) for odd-TLs (such as SM$_{32}$, SM$_{54}$ and SM$_{76}$) and an E$_{u}$ symmetry (IR-active) for even-TLs. However, these modes cannot be observed under back scattering geometry even though being Raman-active. Similarly, only the interlayer breathing modes with the j$^{th}$-lowest (j=1, 3, 5, ...) frequencies (such as LBM$_{NN-1}$, LBM$_{NN-3}$) are Raman-active and can be observed under the VV polarization configuration, while the one with the j$^{th}$-lowest (j=2, 4, 6, ...) frequencies (such as LBM$_{NN-2}$) are all exclusively IR-active modes, with symmetries of $A''_{2}$ for odd-TLs and A$_{2u}$ for even-TLs. However, in our RRS experiments, we not only observed these backscattering forbidden phonon modes (SM$_{32}$, SM$_{54}$ and SM$_{76}$), but also IR-active LBM$_{NN-2}$ modes. We will discuss the physics behind these phenomena in section of discussion.

\begin{figure*}[htb]
\centerline{\includegraphics[width=170mm,clip]{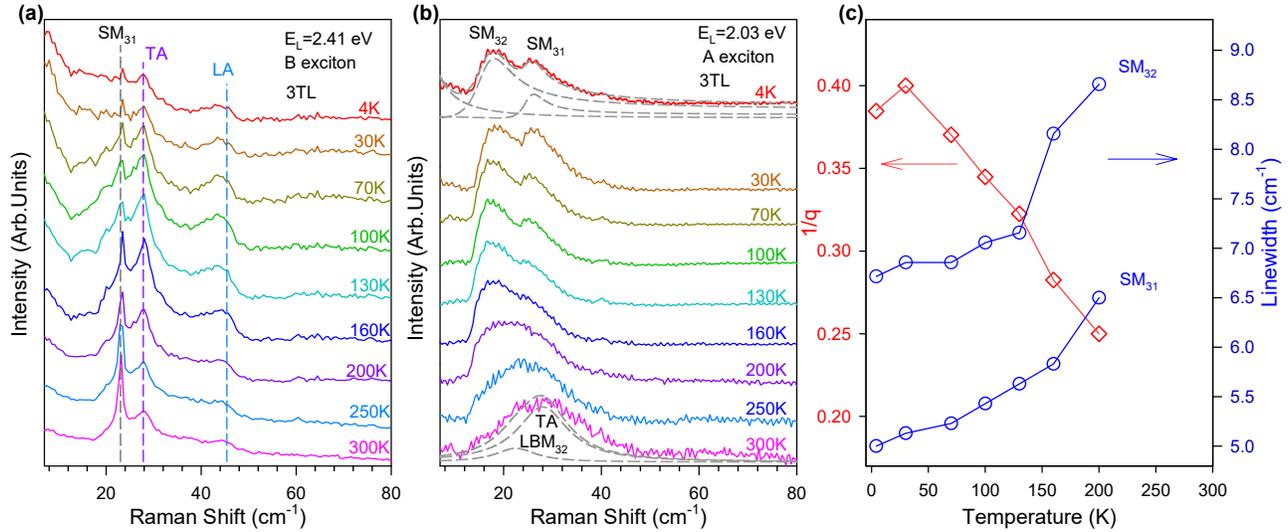}}
\caption{\textbf{Temperature dependence of resonance Raman spectra of 3TL WS$_2$ excited near B and A transitions.} (a-b), Raman spectra of 3TL WS$_2$ excited by 2.41 eV, 2.03 eV lasers, from 4 to 300K, respectively. (c), The fitted parameters $1/q$ and $\gamma$ as a function of temperature.} \label{Fig4}
\end{figure*}

Instead of varying the excitation energies and sample layer numbers, we also conducted the RRS experiments by changing the sample temperature from 4 to 300 K. Figure 4(a) shows the temperature dependence of RRS spectra in 3TL WS$_2$ excited by 2.41 eV laser, which is close to the B exciton transition of 3TL WS$_2$ around 110 K. Both of the TA and LA modes show resonant enhancement behaviors when the B exciton transition is close to the excitation laser energy, but the SM$_{31}$ mode shows non-resonant behavior that the intensity gradually decreases with decreasing temperature. More interesting, as shown in Figure 4(b) spectra excited by 2.03 eV laser, two Fano peak gradually emerge bellow 200 K and become more and more resolvable with decreasing temperature further. Here the laser energy of 2.03 eV is close to the A exciton transition of 3TL WS$_2$ at 30 K. By comparing the LCM predicted frequencies of SM$_{32}$ and SM$_{31}$, we definitely confirm that these two peaks are from the Fano resonance between SM$_{32}$, SM$_{31}$ and electronic continuum state, respectively. When the temperature is higher than 200 K, A exciton transition red shifts to lower energy closed to 1.96 eV and the SM$_{32}$ and SM$_{31}$ Fano resonance profiles are replaced by acoustic phonon TA dominated profile as shown in Figure 4(b). This is consisted with spectra excited by 2.03 eV as shown in Figure 1(a-b). In Figure 4(c), we concluded the fitting parameters of $1/\emph{q}$ and $\gamma$ for SM$_{32}$ and SM$_{31}$ modes as a function of temperature. Since both of $1/\emph{q}$ values SM$_{32}$ and SM$_{32}$ Fano modes are almost the same, and thus we only show the $1/\emph{q}$ value of SM$_{32}$. The value of $1/\emph{q}$ increases with temperature decreases below 30K and then decreases with temperature increases from 30 to 300 K. It indicates that the strongest coupling between phonons and electronic continuum happens at 30 K, where the A exciton transition energy is exactly resonant with excitation laser. The phonon linewidths of ${\gamma}$ are monotonically broadening with temperature increases due to arharmonic effects of phonons and relaxations of excitons\cite{linewidth-prb-1984, RICHTER-SSC-1981}.

Besides observations of forbidden phonon modes in ultra-low frequency range, at low temperature, we also found another type of Raman forbidden but IR-active high frequency modes at A exciton resonance. We conducted the Raman spectra of 2-4TL WS$_2$ excited by 2.03 eV from 4 to 300 K. Figure 5(a) exhibits that $\emph{N}$ Raman peaks are observed in $\emph{N}$TL WS$_2$ (N=2, 3, 4) around $\sim$ 420cm$^{-1}$. These Raman modes become more and more resolvable with decreasing sample temperature, as shown in Figure 5(b). Moreover, as shown in Figure 5(c), these splitting modes show strongly intensity enhancement at A exciton resonant excitation. Based on the peak frequencies and peak numbers, we can definitely assign them as Davydov splitting components of A$'_{1}$-like modes in 2-4TL WS$_2$, respectively. In $\emph{N}$TL 2H MX$_2$ samples, each normal mode of 1TL will turn into $\emph{N}$ Davydov splitting modes, keeping the intra-TL displacement while varying the phase difference (0$^\circ$ or 180$^\circ$) between adjacent TLs\cite{ mote2-prb-2016}. These Davydov components are either Raman or IR-actived. Their total number is equal to the layer number. Therefore, all the Raman (R) and infrared (IR) active Davydov components of A$'_{1}$-like modes are observed in our experiments for 2-4TL, as labeled in Figure 5(a). Although the Davydov splitting has been observed in few-layer MoTe$_2$\cite{mote2-prb-2016,Froehlicher-nanoletter-2015}, MoSe$_2$\cite{Tonndorf-oe-2013,kim-nanoletter-2016}, and twisted graphene\cite{Wu-nc-2014}, the IR active components are still rarely reported. Recently, Staiger $\emph{et al}$ reported an experimental observation of IR-active and Raman-active Davydov components in few-layer WS$_2$\cite{Staiger-prb-2015} by using resonant Raman scattering at room temperature, but the spectra are not resolved very well and it is not clear why IR-active phonons can be observed in Raman scattering. Here, by using the RRS at low temperature, we can clearly resolve all of Davydov components of A$'_{1}$-like modes in 2-4TL WS$_2$, including the Raman active A$'_{1}/A_{1g}$ and IR active A$_{2u}/A''_2$ modes.

\begin{figure*}[htb]
\centerline{\includegraphics[width=170mm,clip]{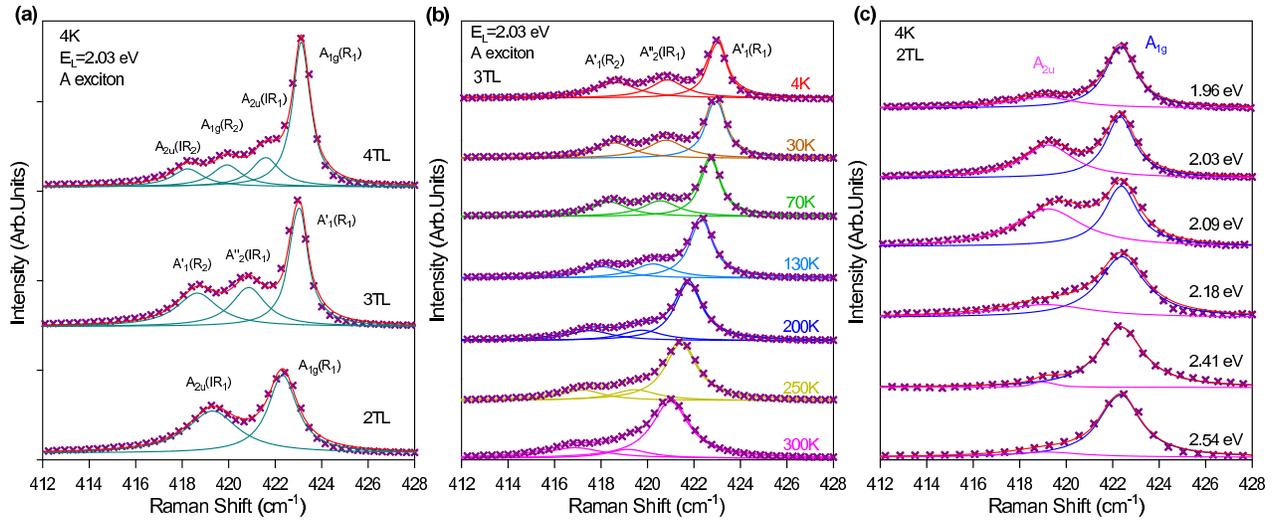}}
\caption{\textbf{Temperature dependent Davydov components of A$'_1$-like modes} (a), Raman spectra of A$'_{1}$-like Davydov components in 2-4TL WS$_2$ excited by 2.03 eV laser at 4 K. (b), Temperature dependent Raman spectra of A$'_{1}$-like Davydov components in 3TL WS$_2$ excited by the 2.03 eV laser. (c), Raman spectra of A$_{1g}/A_{2u}$ modes in 2TL WS$_2$ with different excitation energies at 4K. The solid lines refer to the Lorentzian fitting results. The cross symbols are the experimental results. All the spectra are normalized to main peak and are offset for clarity.} \label{Fig5}
\end{figure*}

\subsection*{Discussions}
So far, several forbidden phonon modes have been observed in RRS spectra of few-layer WS$_2$, including IR-active phonons (low frequency LBM$_{NN-2}$, and high requency A$_{2u}/A''_2$ modes) and Raman-active but backscattering forbidden phonons (SM$_{32}$, SM$_{54}$ and SM$_{76}$). Besides, the LA mode shows a Fano lineshape with negative $\emph{q}$ in both of A and B exciton resonance, and the peak intensity is much stronger in B exction resonance than in A exciton resonance; the SM$_{NN-1}$ and SM$_{31}$ show the positive $\emph{q}$ Fano lineshape only under A exciton resonance. In RRS, the observation of backscattering forbidden phonons can be explained by Fr$\ddot{o}$hlich mechanism of exction-phonon interaction\cite{PhysRevLett-26-86}. In RRS, Fr$\ddot{o}$hlich interaction between excitons and phonons leads to the forbidden scattering intensity is proportional to $(\emph{k}_{p}a)^2$, where $\emph{k}_{p}$ is wave vector of phonon and $\emph{a}$ is a Bohr radius of the exciton. Thus, forbidden phonons are observed only if $\emph{a}$ is much larger than the lattice constant. In this case, forbidden phonons in backscattering get the largest scattering cross-section because $\emph{k}_{p}$ is the largest in backscattering. These conditions can be satisfied in WS$_{2}$ because the excitons in WS$_2$ are large-radius Wannier excitons\cite{Dark-exciton-nature-2014} and they can couple to phonons via the intraband Fr$\ddot{o}$hlich interaction.

The observation of IR-active phonon in our RRS experiments can be explained by analyzing the RRS process. The first-order resonant Raman scattering probability is given by\cite{ cu2o-prl-1974,PhysRevLett-3141-1973}:
\begin{equation}
\begin{split}
&P\propto\mid\frac{\langle{i}|H_{eR}(\omega_s)| {m'}\rangle\langle{m'}| H_{eL}|{m}\rangle\langle{m}|H_{eR}(\omega_i)| {i}\rangle}{(E_{m'}-\hbar\omega_i-i\gamma_{m'})( E_{m}-\hbar\omega_s-i\gamma_m)} \\
&+ \frac{\langle{i}|H_{eR}(\omega_i)| {m}\rangle\langle{m}| H_{eL}|{m'}\rangle\langle{m'}|H_{eR}(\omega_s)| {i}\rangle}{(E_{m}-\hbar\omega_i-i\gamma_{m})( E_{m'}-\hbar\omega_s-i\gamma_{m'})} \mid^2\\
\end{split}
\end{equation}
where ${\omega_i}$, ${\omega_{ph}}$ and $\omega_s = \omega_i \pm \omega_{ph}$ is the incident photon, phonon and scattering photon frequencies, and $m'$ and $m$ are intermediate exciton state with energy E$_{m'}$ and E$_{m'}$ and widths $\gamma_{m'}$, and ${\gamma_m}$, respectively. ${H_{eL}}$ denotes the exciton-lattice interaction; ${H_{eR}}$ denotes the exciton-radiation-field interaction.

In few-layer WS$_2$, the spin-orbit splitting of conduction band leads to both of 1s A and B exciton split into dark and bright exctonic fine structures, respectively. Bright state has odd parity and is dipole (E1) transition allowed; dark state has even parity and is dipole forbidden but electric quadrupole (E2) or magnetic dipole (M1) allowed\cite{ZhangXX-prl-2015,Splitting-PRB-2016}. Beside excitons, phonons also have definite parity based on group theory: IR-active phonons have odd parity but Raman-active phonons have even parity\cite{ DMS-1984}. Since exction-phonon interaction $H_{eL}$ has the same symmetry with scattered phonons, the parity of $H_{eL}$ is the same as the parity of phonons. In order to get a nonzero transition in equation (2), the parity of $H_{eL}$ has to be the same as parity of $H_{eR}{(\omega_i)}\otimes{H_{eR}{(\omega_s)}}$ \cite{ cu2o-prl-1974,PhysRevLett-3141-1973}. In non-resonant or bright exciton resonant condition, since both of $H_{eR}{(\omega_i)}$ and $H_{eR}{(\omega_s)}$ are electronic-dipole allowed E1 transitions with odd parity, and thus the product of ${H_{eR}{(\omega_i)}}\otimes{H_{eR}{(\omega_s)}}$ is even parity. Therefore, a nonzero transition in equation (2) is required that the $H_{eL}$ (and thus phonons) must be even parity (Raman-active). This is equivalent to the classical statement that a phonon is Raman active in first order Raman scattering only when its parity is even. When the dark exciton involved in the resonant Raman transition, for example, the incident resonance case, the ${H_{eR}(\omega_i)}$ is E2 or M1 transition with even parity and the off-resonant out-going transition ${H_{eR}(\omega_s)}$ is the usual E1 transition with odd parity, and thus the product of $H_{eR}{(\omega_i)}\otimes{H_{eR}{(\omega_s)}}$ is odd parity. Therefore, the phonons related to $H_{eL}$ must be odd parity (IR-active). It has the same result in out-going resonant transition. In conclusion, when the incident or scattering photons resonate with odd-parity dark-excitons, odd parity IR-active phonons can be observed based on the parity selection rule of Raman scattering. This mechanism, therefore, explains why the odd parity IR-active modes can be involved in RRS in few-layer WS$_2$.

Another factor to affect the Raman intensity in equation (2) is damping constant of intermediate exciton states. Normally one expects the E2 and M1 transition matrix elements in Eq. (2) to be several orders of magnitude weaker than the E1 transition matrix elements. Generally, such forbidden IR-active phonon modes should be too weak to be observed. However, the small optical matrix elements result in small radiative decay probabilities and hence small damping constants $\gamma_{m', m}$ of intermediate excitons. At resonance, the small damping constant of dark excitons in the denominators in Eq. (2) over-compensates the small matrix elements in the numerator, because the damping constants are proportional to the square of the optical matrix elements. Since damping constants are inverse proportional to the lifetime, this means dark excitons have much longer life-time than bright excitons. This is particularly true for 1s excitons in TMDs. For example, the lifetime of dark exciton is in the order of from nanosecond in monolayer TMDs\cite{ZhangXX-prl-2015,trion-NC-2016} to microsecond in 2D heterostructures\cite{Xiong-arXiv-2017}, which are much longer than that of the bright exciton in an order of picosecond in TMDs. Besides, we should note that the IR-active modes and Fano lineshape SMs only are observed under A exciton resonance but vanished under B exciton resonance. This is because the oscillator strength of A and B dark excitons are very different. Although the oscillator strength of A exciton dark state is quite small, but it is not strictly zero because its single particle state can be slightly mixed with remote bands\cite{XD-prl-2012,Splitting-PRB-2016}. However, for dark B-exciton state, its symmetry can be represented as $\Gamma_{6}$. Therefore, the oscillator strength of dark B-exciton is strictly equal to zero\cite{Splitting-PRB-2016}. This is the reason why we could almost not observe IR-active modes and Fano lineshape SMs under B exciton transitions. Comparing the other techniques used to detect the excitonic dark states in few-layer TMDs\cite{Dark-exciton-nature-2014,ZhangXX-prl-2015,trion-NC-2016}, our RRS technique can simultaneously detect the forbidden phonons and dark exciton states.

Now let's discuss physical mechanisms behind the Fano resonance of SM$_{NN-1}$, SM$_{31}$ and $LA$ modes. Figure S5 in supplementary information presents the schematic diagram of dark-bright energy splitting of the 1s A excitons in WS$_2$ systems at K point\cite{Splitting-PRB-2016}. We should note linewidthes of bright and dark excitons are much broader than that of phonons, and even larger than the phonon energies involved in Fano resonance. In this case, the excitons act like a continuum state and phonon is discrete state. When the incident laser energy resonates with exciton transition, the laser-induced excited state will be populated and interferences with discrete phonons. Depending on the exciton-phonon coupling strength, some of phonons will show Fano resonance. There are two types of interferences between discrete phonon and continuum exciton states: one type is interband transition involved $\emph{k}$$\simeq$0 phonon; the other one is intraband or interband transition involved $\emph{k}$$\neq$0, where intraband transition is spin conservation and interband transition is accompany by spin-flip scattering. From our experiments, we found that when the excitation energy (for example, 1.92 and 1.96 eV) is slightly lower than the bright A exciton transitions at room temperature, the Fano peaks of SM$_{NN-1}$ are much stronger than that excited by 2.03 eV laser, which is slightly higher than the A exciton transitions, as shown in Figure 3(b,c) and supporting Figure S6. This result indicates that the exciton continuum must be below the bright exciton state. This proposal also can be further confirmed by measuring the Raman spectra from 4 to 300K with excitation energy of 2.03 eV. When temperature is above 200K, the Fano-type SMs almost vanished because the excitation energy is slightly larger than the A exciton transitions at room temperature. According to our experiments, the Fano lineshape of SMs appeared at a range from 1.92 eV to 1.96 eV at room temperature. Therefore, we can estimate dark-bright splitting of 1s A exciton is around -40 meV, which has the same sign but the splitting energy is larger than recent calculations of -20 meV\cite{NC-darkexciton-2017} and -11 meV\cite{Splitting-PRB-2016}. For B exciton, we can't identify the bright-dark splitting based our results.

\subsection*{CONCLUSIONS}
In conclusion, by studying the exctonic RRS spectra in few-layer and bulk WS$_2$ excited by multiple wavelength laser lines, and at different sample temperatures from 4 to 300 K, we have observed several forbidden phonon modes that is unobservable in ordinary Raman scattering experiments. By analyzing the parity selection rule of Raman scattering and Fano resonance of SM and LA modes, we can not only explain our results very well, but also definitely determine the bright-dark fine structure of 1s A exciton. Our results not only provide a better understanding of exciton-phonon interactions, but also show alternative technique to simultaneously probe the dark excitons and forbidden phonons in few-layer TMDs. Besides, our experiment also presents an opportunity to study the strong many-body effects between light, excitons and phonons, and thus help the development of 2D materials in optoelectronic applications.
\\
\subsection*{Methods}

\subsubsection*{\bf Sample preparation.} The samples were prepared from a bulk WS$_2$ crystals onto a 90 nm SiO$_2/$Si substrate and SiO$_2$ substrate by using the micromechanical exfoliation technique.

\subsubsection*{\bf Raman and Reflectance contrast $\Delta{R/R}$ measurements.} Raman measurements were undertaken in backscattering geometry with a Jobin-Yvon HR800 system equipped with a liquid-nitrogen-cooled charge-coupled detector. The spectra were collected with a 100 x objective lens (NA=0.9) and an 1800 lines mm$^{-1}$ grating at room temperature, while a 50 x long-working-distance objective lens (NA=0.5) was used at low temperature measurements. The excitation laser (E$_L$) lines of 2.71 eV, 2.54 eV and 2.41 eV are from an Ar$^{+}$ laser; the laser lines of 2.09 eV, 2.03 eV and 1.96 eV are from a He-Ne laser; the laser lines of 2.34 eV, 2.18 eV, 1.92 eV and 1.83 eV are from a Kr$^{+}$ laser, the laser line of 2.81 eV is from a He-Cd laser. The ultralow-frequency Raman spectra were obtained down to $\pm$5 cm$^{-1}$ by combining three volume Bragg grating filters into the Raman system to efficiently suppress the Rayleigh signal. In order to avoid the laser heating effect to the samples, the laser power was kept below a maximum of 0.3 mW. The Montana cryostat system was employed to cool the samples down to 4K under a vacuum of 0.4 mTorr. The reflectance contrast $\Delta{R/R}$ were undertaken with a 100x objective lens (NA=0.9) and an 100 lines mm$^{-1}$ grating with white light at room temperature.

\subsubsection*{\bf Acknowledgements}
\noindent J.Z. and P.T. acknowledge support from National Basic Research Program of China (grant no. 2016YFA0301200), NSFC (11574305, 51527901, 11474277, 11434010 and 11225421) and and LU JIAXI International team program supported by the K.C. Wong Education Foundation and the Chinese Academy of Sciences. J. Z. also acknowledges support from National Young 1000 Talent Plan of China; Q.X. acknowledges Singapore National Research Foundation via NRF Investigatorship Award (NRF-NRFI2015-03), and the Singapore Ministry of Education via AcRF Tier 2 grant (MOE2012-T2-2-086 and MOE2015-T2-1-047).

\vspace*{5mm}
\noindent {\bf \large Author contribution}

\noindent J.Z. conceived the ideas; Q.T., P.T., Y.Z., and J. Z. designed the experiments. Q.T., Y.S. prepared the samples. Q.T., X.L. and Y.Z. performed experiments; Q.T., P.T. and Z.J. analyzed the data; Q.T., and Z.J. wrote the manuscript with input from all authors. J. Z., P.T. and Q.X. supervised the projects.

\vspace*{5mm}
\noindent {\bf \large Additional information}

\noindent {\bf Supplementary information} is available in the online version of the paper. Reprints and permissions information is available online at www.nature.com/reprints. Correspondence and requests for materials should be addressed to P.T.( Email: phtan@semi.ac.cn) and J.Z.(Email: zhangjwill@semi.ac.cn).
\\
\noindent {\bf Competing interests:} The authors declare no competing financial interest.

\vspace*{40mm}

\subsection*{\bf \large Supplementary Information}
\begin{figure*}[htb]
\centerline{\includegraphics[width=150mm,clip]{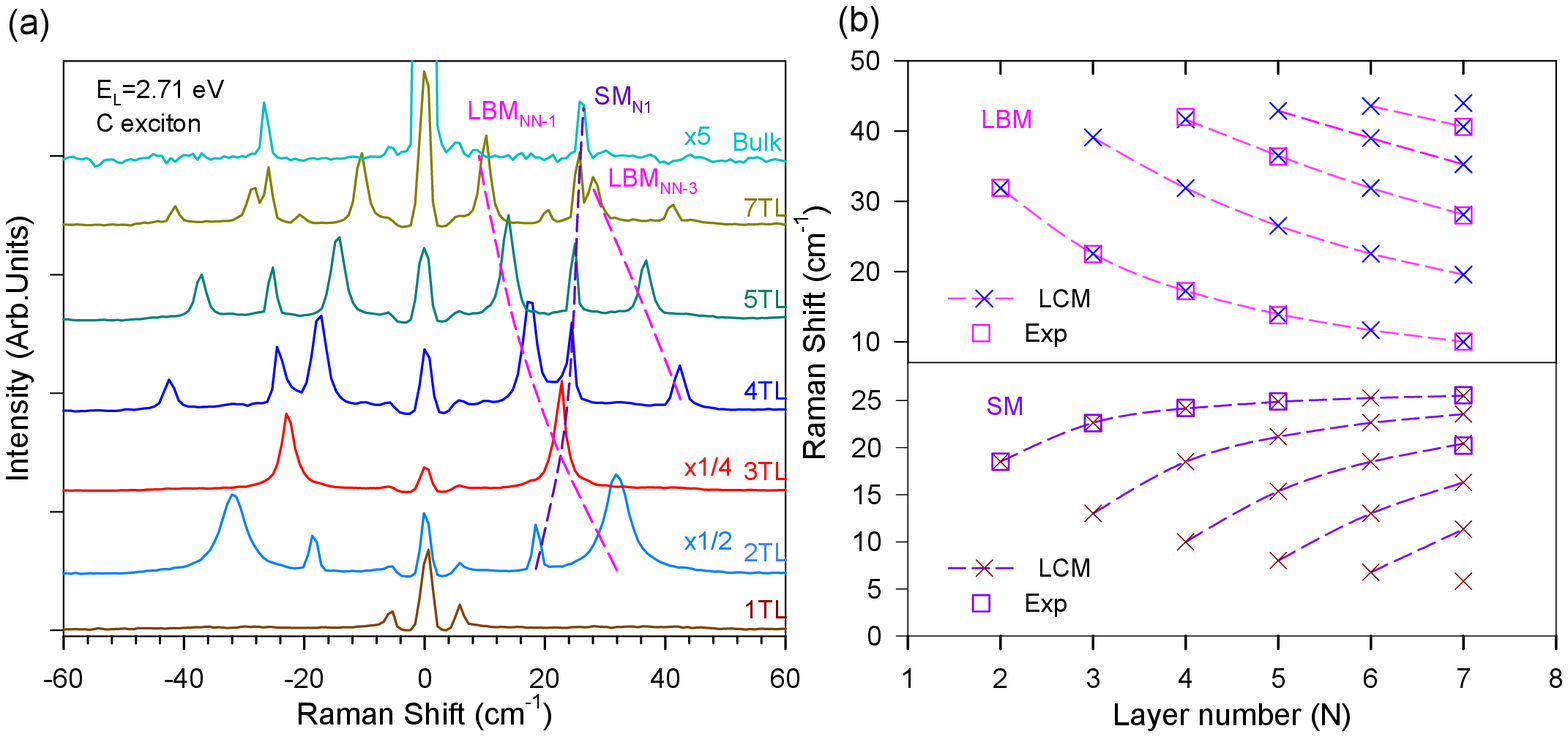}}
\caption{\textbf{Figure S1} (a) Ultralow frequency Raman spectra of 1-5L, 7TL, and bulk WS$_2$ excited by 2.71 eV laser. (b) Experimental and calculated frequencies of interlayer shear modes (SM) and layer-breathing modes (LBM) in 1-7TL WS$_2$. Here the calculation is based on linear chain model (LCM)}\cite{Zhang-prb-2013} \label{Figure S1}
\end{figure*}

\begin{figure*}[htb]
\centerline{\includegraphics[width=150mm,clip]{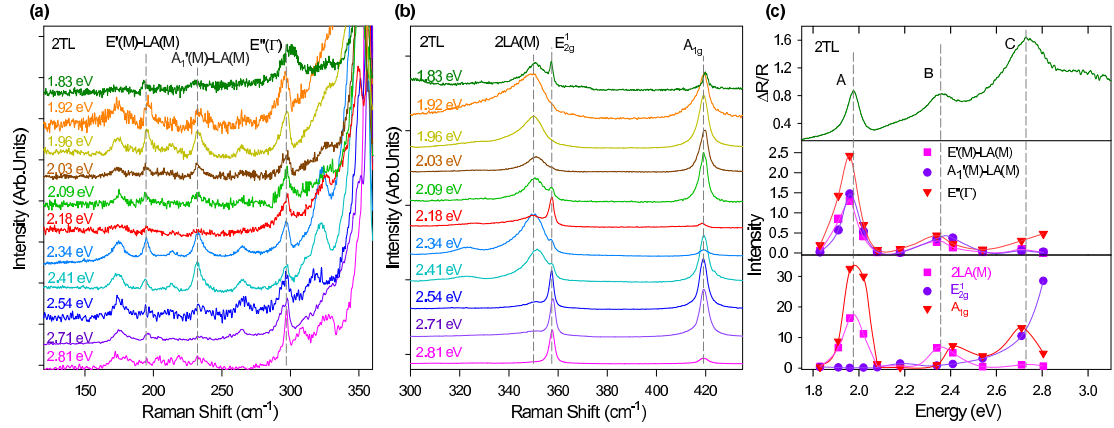}}
\caption{\textbf{Figure S2}(a-b) Raman spectra of 2TL WS$_2$ with multiple excitation wavelengths. (c) Reflectance contrast $\Delta$R/R spectra for 2TL WS$_2$ and Raman excitation profiles of the A$_{1g}$, E$^{1}_{2g}$ and second order Raman modes in (a-b). The solid lines are used to guide to the eyes.} \label{Figure S2}
\end{figure*}

\begin{figure*}[htb]
\centerline{\includegraphics[width=150mm,clip]{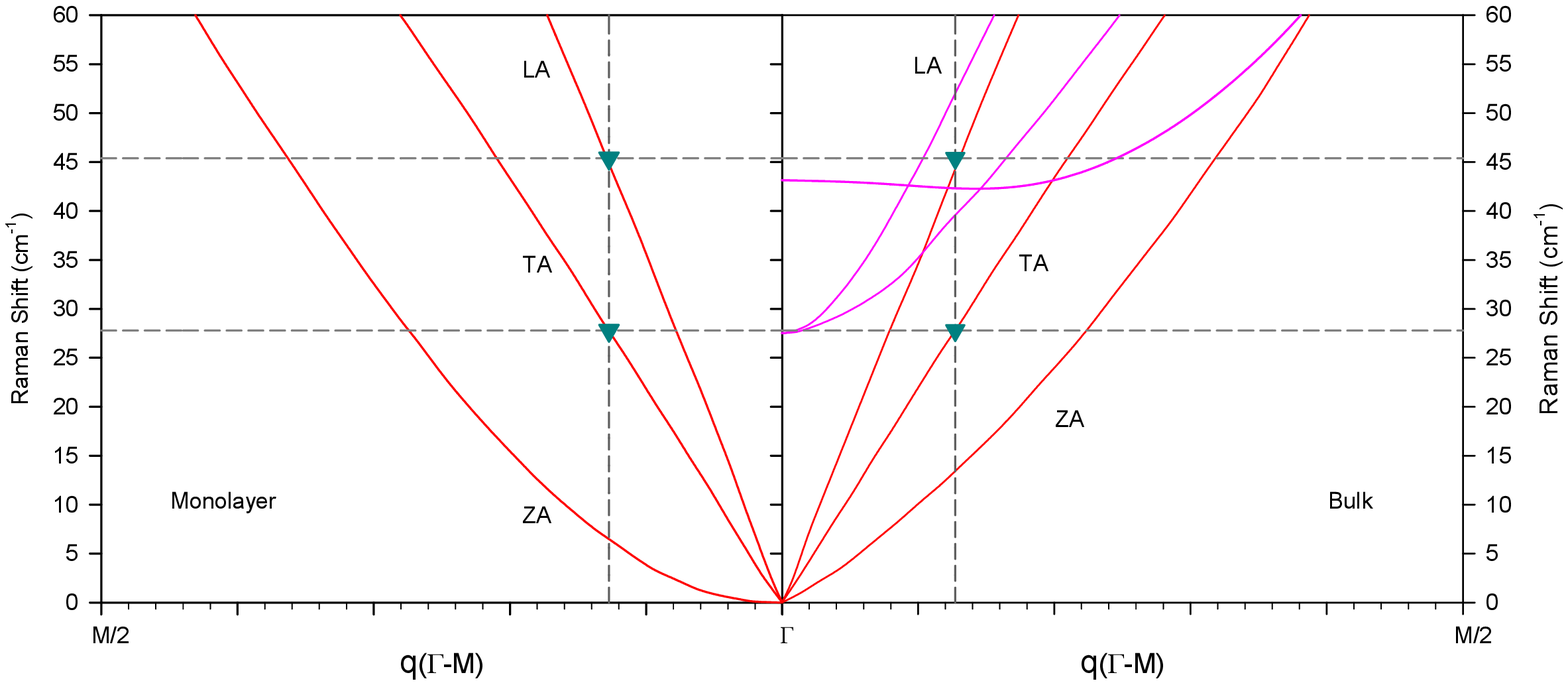}}
\caption{\textbf{Figure S3} The dispersion curves below 60 cm$^{-1}$ close to Brillouin zone $\Gamma$ point in monolayer and bulk WS$_2$\cite{Phonondispersion-2011}. The horizontal short dash lines represent the frequencies of observed modes that independent on excitation energies and sample layers. The triangle symbols at the cross point between the phonon dispersion and horizontal short dash represent corresponding phonons selected by finite $\emph{k}$. Pink curves are phonon branches related with shear and breathing modes. LA, TA and ZA refers to longitudinal acoustic (LA), transversal acoustic (TA) and out-of-plane acoustic (ZA) phonons, respectively.} \label{Figure S3}
\end{figure*}
\begin{figure*}[htb]
\centerline{\includegraphics[width=150mm,clip]{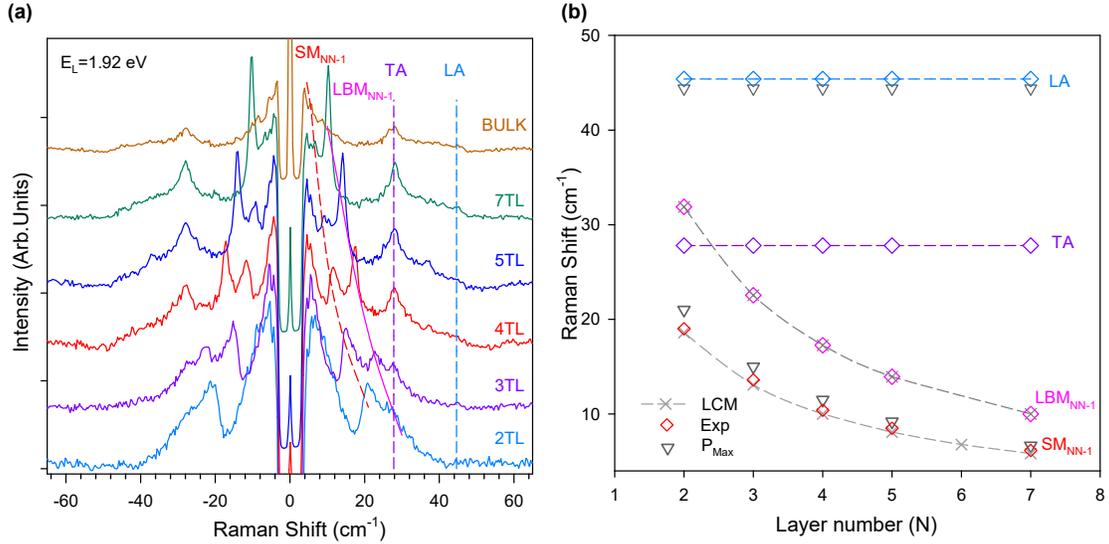}}
\caption{\textbf{Figure S4} (a) The RRS spectra of 1-5TL, 7TL and bulk WS$_2$ crystals excited by laser of 1.92 eV (647 nm). (b) The layer number dependence of phonon frequencies for experimental (diamonds) and calculated (cross) results. LCM represents the results based on the liner chain model\cite{Zhang-prb-2013}.} \label{Figure S4}
\end{figure*}

\begin{figure*}[htb]
\centerline{\includegraphics[width=70mm,clip]{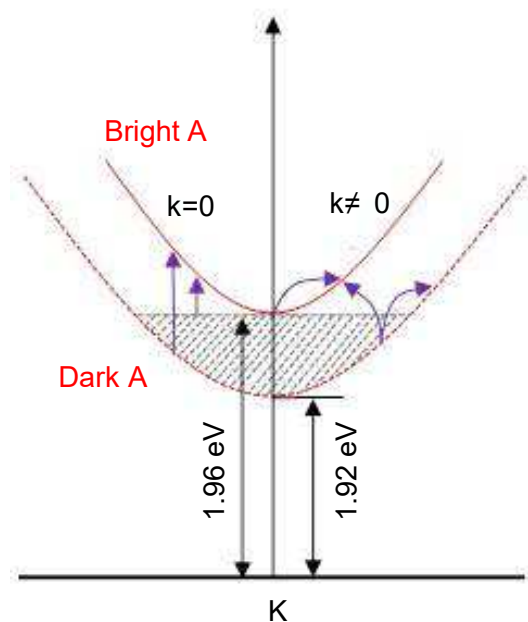}}
\caption{\textbf{Figure S5} A schematic diagram of bright-dark splitting and phonon transition ways involved in Fano interferences for 1s A exciton.} \label{Figure S5}
\end{figure*}

\begin{figure*}[htb]
\centerline{\includegraphics[width=150mm,clip]{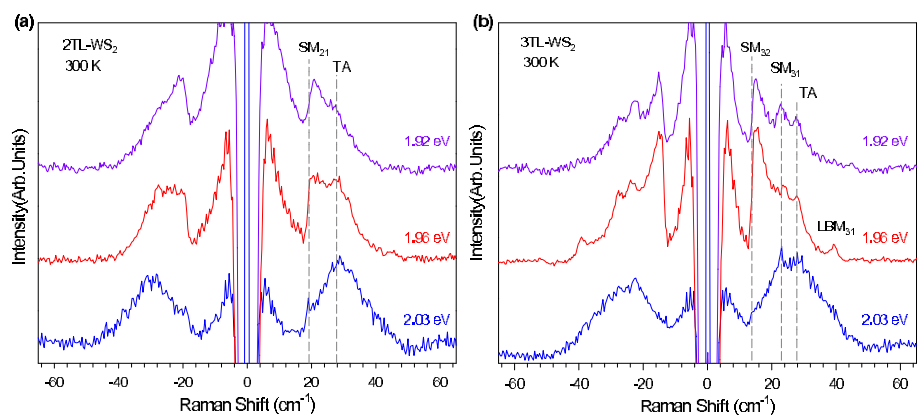}}
\caption{\textbf{Figure S6} Raman spectra of (a) 2TL and (b) 3TL WS$_2$ excited by 1.92 eV, 1.96 eV and 2.03 eV lasers, respectively. The intensities are normalized to the strongest mode.} \label{Figure S6}
\end{figure*}


\vspace*{20mm}
\noindent {\bf \large References}
\begin{thebibliography}{10}
\expandafter\ifx\csname url\endcsname\relax
  \def\url#1{\texttt{#1}}\fi
\expandafter\ifx\csname urlprefix\endcsname\relax\def\urlprefix{URL }\fi
\providecommand{\bibinfo}[2]{#2}
\providecommand{\eprint}[2][]{\url{#2}}

\bibitem{ZJ-Nature-2013}
\bibinfo{author}{Zhang, J.}, \bibinfo{author}{Li, D.}, \bibinfo{author}{Chen,
  R.} \& \bibinfo{author}{Xiong, Q.}
\newblock \bibinfo{title}{Laser cooling of a semiconductor by 40 kelvin}.
\newblock \emph{\bibinfo{journal}{Nature}} \textbf{\bibinfo{volume}{493}},
  \bibinfo{pages}{504--508} (\bibinfo{year}{2013}).

\bibitem{ZJ-NP-2016}
\bibinfo{author}{Zhang, J.}, \bibinfo{author}{Zhang, Q.},
  \bibinfo{author}{Wang, X.}, \bibinfo{author}{Kwek, L.~C.} \&
  \bibinfo{author}{Xiong, Q.}
\newblock \bibinfo{title}{Resolved-sideband raman cooling of an optical phonon
  in semiconductor materials}.
\newblock \emph{\bibinfo{journal}{Nat Photon}} \textbf{\bibinfo{volume}{10}},
  \bibinfo{pages}{600--605} (\bibinfo{year}{2016}).

\bibitem{polaritons-prb-1979}
\bibinfo{author}{Koteles, E.~S.} \& \bibinfo{author}{Winterling, G.}
\newblock \bibinfo{title}{Resonant scattering of exciton polaritons by {LO} and
  acoustic phonons}.
\newblock \emph{\bibinfo{journal}{Phys. Rev. B}} \textbf{\bibinfo{volume}{20}},
  \bibinfo{pages}{628--637} (\bibinfo{year}{1979}).

\bibitem{CdS-PRL}
\bibinfo{author}{Leite, R. C.~C.}, \bibinfo{author}{Scott, J.~F.} \&
  \bibinfo{author}{Damen, T.~C.}
\newblock \bibinfo{title}{{Multiple-Phonon Resonant Raman Scattering in CdS}}.
\newblock \emph{\bibinfo{journal}{Phys. Rev. Lett.}}
  \textbf{\bibinfo{volume}{22}}, \bibinfo{pages}{780--782}
  (\bibinfo{year}{1969}).

\bibitem{PhysRevB-85115317-2012}
\bibinfo{author}{Kaasbjerg, K.}, \bibinfo{author}{Thygesen, K.~S.} \&
  \bibinfo{author}{Jacobsen, K.~W.}
\newblock \bibinfo{title}{{Phonon-limited mobility in $n$-type single-layer
  MoS${}_{2}$ from first principles}}.
\newblock \emph{\bibinfo{journal}{Phys. Rev. B}} \textbf{\bibinfo{volume}{85}},
  \bibinfo{pages}{115317} (\bibinfo{year}{2012}).

\bibitem{Carvalho-prl-2015}
\bibinfo{author}{Carvalho, B.~R.}, \bibinfo{author}{Malard, L.~M.},
  \bibinfo{author}{Alves, J.~M.}, \bibinfo{author}{Fantini, C.} \&
  \bibinfo{author}{Pimenta, M.~A.}
\newblock \bibinfo{title}{{Symmetry-Dependent Exciton-Phonon Coupling in 2D and
  Bulk MoS$_2$ Observed by Resonance Raman Scattering}}.
\newblock \emph{\bibinfo{journal}{Phys. Rev. Lett.}}
  \textbf{\bibinfo{volume}{114}}, \bibinfo{pages}{136403}
  (\bibinfo{year}{2015}).

\bibitem{um-nm-2014}
\bibinfo{author}{Ugeda, M.~M.} \emph{et~al.}
\newblock \bibinfo{title}{Giant bandgap renormalization and excitonic effects
  in a monolayer transition metal dichalcogenide semiconductor}.
\newblock \emph{\bibinfo{journal}{Nat Mater}} \textbf{\bibinfo{volume}{13}},
  \bibinfo{pages}{1091--1095} (\bibinfo{year}{2014}).

\bibitem{Qiu-PRL-2013}
\bibinfo{author}{Qiu, D.~Y.}, \bibinfo{author}{da~Jornada, F.~H.} \&
  \bibinfo{author}{Louie, S.~G.}
\newblock \bibinfo{title}{{Optical Spectrum of MoS$_2$: Many-Body Effects and
  Diversity of Exciton States}}.
\newblock \emph{\bibinfo{journal}{Phys. Rev. Lett.}}
  \textbf{\bibinfo{volume}{111}}, \bibinfo{pages}{216805}
  (\bibinfo{year}{2013}).

\bibitem{cuixiaodong-serp-2015}
\bibinfo{title}{Exciton binding energy of monolayer ws$_2$} .

\bibitem{Xu-NPRiew-2016}
\bibinfo{author}{Xu, X.~D.}, \bibinfo{author}{Yao, W.}, \bibinfo{author}{Xiao,
  D.} \& \bibinfo{author}{Heinz, T.~F.}
\newblock \bibinfo{title}{Spin and pseudospins in layered transition metal
  dichalcogenides}.
\newblock \emph{\bibinfo{journal}{Nat Phys}} \textbf{\bibinfo{volume}{10}},
  \bibinfo{pages}{343--350} (\bibinfo{year}{2014}).

\bibitem{Cao-NC-2012}
\bibinfo{author}{Cao, T.} \emph{et~al.}
\newblock \bibinfo{title}{Valley-selective circular dichroism of monolayer
  molybdenum disulphide}.
\newblock \emph{\bibinfo{journal}{Nature communications}}
  \textbf{\bibinfo{volume}{3}}, \bibinfo{pages}{887} (\bibinfo{year}{2012}).

\bibitem{cuixiaodong-PNAS-2014}
\bibinfo{author}{Zhu, B.}, \bibinfo{author}{Zeng, H.}, \bibinfo{author}{Dai,
  J.}, \bibinfo{author}{Gong, Z.} \& \bibinfo{author}{Cui, X.}
\newblock \bibinfo{title}{Anomalously robust valley polarization and valley
  coherence in bilayer {WS$_2$}}.
\newblock \emph{\bibinfo{journal}{Proceedings of the National Academy of
  Sciences}} \textbf{\bibinfo{volume}{111}}, \bibinfo{pages}{11606--11611}
  (\bibinfo{year}{2014}).

\bibitem{cuixiaodong-review-2015}
\bibinfo{author}{Zeng, H.} \& \bibinfo{author}{Cui, X.}
\newblock \bibinfo{title}{{An optical spectroscopic study on two-dimensional
  group-VI transition metal dichalcogenides}}.
\newblock \emph{\bibinfo{journal}{{Chem. Soc. Rev.}}}
  \textbf{\bibinfo{volume}{44}}, \bibinfo{pages}{2629--2642}
  (\bibinfo{year}{2015}).

\bibitem{Dark-exciton-nature-2014}
\bibinfo{author}{Ye, Z.} \emph{et~al.}
\newblock \bibinfo{title}{Probing excitonic dark states in single-layer
  tungsten disulphide}.
\newblock \emph{\bibinfo{journal}{Nature}} \textbf{\bibinfo{volume}{513}},
  \bibinfo{pages}{214--218} (\bibinfo{year}{2014}).

\bibitem{NP-xuxiaodong-2016}
\bibinfo{author}{Jones, A.~M.} \emph{et~al.}
\newblock \bibinfo{title}{Excitonic luminescence upconversion in a
  two-dimensional semiconductor}.
\newblock \emph{\bibinfo{journal}{Nat Phys}} \textbf{\bibinfo{volume}{12}},
  \bibinfo{pages}{323--327} (\bibinfo{year}{2016}).

\bibitem{mm-arXiv-2017}
\bibinfo{author}{Manca, M.} \emph{et~al.}
\newblock \bibinfo{title}{Enabling valley selective exciton scattering in
  monolayer {WSe$_2$} through upconversion}.
\newblock \emph{\bibinfo{journal}{arXiv:1701.05800}}  (\bibinfo{year}{2017}).

\bibitem{NM-LowT-2017}
\bibinfo{author}{Low, T.} \emph{et~al.}
\newblock \bibinfo{title}{Polaritons in layered two-dimensional materials}.
\newblock \emph{\bibinfo{journal}{Nat Mater}} \textbf{\bibinfo{volume}{16}},
  \bibinfo{pages}{182--194} (\bibinfo{year}{2017}).

\bibitem{Zhang-Xin-CSR-2015}
\bibinfo{author}{Zhang, X.} \emph{et~al.}
\newblock \bibinfo{title}{Phonon and raman scattering of two-dimensional
  transition metal dichalcogenides from monolayer, multilayer to bulk
  material}.
\newblock \emph{\bibinfo{journal}{Chemical Society Reviews}}
  \textbf{\bibinfo{volume}{44}}, \bibinfo{pages}{2757--85}
  (\bibinfo{year}{2015}).

\bibitem{Lux-nanoresarch-2016}
\bibinfo{author}{Lu, X.}, \bibinfo{author}{Luo, X.}, \bibinfo{author}{Zhang,
  J.}, \bibinfo{author}{Quek, S.~Y.} \& \bibinfo{author}{Xiong, Q.}
\newblock \bibinfo{title}{{Lattice vibrations and Raman scattering in
  two-dimensional layered materials beyond graphene}}.
\newblock \emph{\bibinfo{journal}{Nano Research}} \textbf{\bibinfo{volume}{9}},
  \bibinfo{pages}{3559--3597} (\bibinfo{year}{2016}).

\bibitem{Tan-nm-2012}
\bibinfo{author}{Tan, P.~H.} \emph{et~al.}
\newblock \bibinfo{title}{The shear mode of multilayer graphene}.
\newblock \emph{\bibinfo{journal}{Nature Materials}}
  \textbf{\bibinfo{volume}{11}}, \bibinfo{pages}{294--300}
  (\bibinfo{year}{2012}).

\bibitem{Zhao-nanoletter-2013}
\bibinfo{author}{Zhao, Y.} \emph{et~al.}
\newblock \bibinfo{title}{Interlayer breathing and shear modes in few-trilayer
  {MoS$_2$ and WSe$_2$}}.
\newblock \emph{\bibinfo{journal}{Nano letters}} \textbf{\bibinfo{volume}{13}},
  \bibinfo{pages}{1007--1015} (\bibinfo{year}{2013}).

\bibitem{Sun-PRL-2013}
\bibinfo{author}{Sun, L.~F.} \emph{et~al.}
\newblock \bibinfo{title}{{Spin-Orbit Splitting in Single-Layer MoS$_2$
  Revealed by Triply Resonant Raman Scattering}}.
\newblock \emph{\bibinfo{journal}{Phys. Rev. Lett.}}
  \textbf{\bibinfo{volume}{111}}, \bibinfo{pages}{126801}
  (\bibinfo{year}{2013}).

\bibitem{del-corro-nanolett-2014}
\bibinfo{author}{del Corro, E.} \emph{et~al.}
\newblock \bibinfo{title}{{Excited Excitonic States in 1L, 2L, 3L, and Bulk
  WSe2 Observed by Resonant Raman Spectroscopy}}.
\newblock \emph{\bibinfo{journal}{ACS Nano}} \textbf{\bibinfo{volume}{8}},
  \bibinfo{pages}{9629--9635} (\bibinfo{year}{2014}).

\bibitem{MAA-PRB-2014}
\bibinfo{author}{Mitioglu, A.~A.} \emph{et~al.}
\newblock \bibinfo{title}{{Second-order resonant Raman scattering in
  single-layer tungsten disulfide ${\mathrm{WS}}_{2}$}}.
\newblock \emph{\bibinfo{journal}{Phys. Rev. B}} \textbf{\bibinfo{volume}{89}},
  \bibinfo{pages}{245442} (\bibinfo{year}{2014}).

\bibitem{mote2-prb-2016}
\bibinfo{author}{Song, Q.~J.} \emph{et~al.}
\newblock \bibinfo{title}{{Physical origin of Davydov splitting and resonant
  Raman spectroscopy of Davydov components in multilayer
  ${\mathrm{MoTe}}_{2}$}}.
\newblock \emph{\bibinfo{journal}{Phys. Rev. B}} \textbf{\bibinfo{volume}{93}},
  \bibinfo{pages}{115409} (\bibinfo{year}{2016}).

\bibitem{MoSe2-prb-2016}
\bibinfo{author}{Soubelet, P.}, \bibinfo{author}{Bruchhausen, A.~E.},
  \bibinfo{author}{Fainstein, A.}, \bibinfo{author}{Nogajewski, K.} \&
  \bibinfo{author}{Faugeras, C.}
\newblock \bibinfo{title}{{Resonance effects in the Raman scattering of
  monolayer and few-layer ${\mathrm{MoSe}}_{2}$}}.
\newblock \emph{\bibinfo{journal}{Phys. Rev. B}} \textbf{\bibinfo{volume}{93}},
  \bibinfo{pages}{155407} (\bibinfo{year}{2016}).

\bibitem{Wu-nc-2014}
\bibinfo{author}{Wu, J.-B.} \emph{et~al.}
\newblock \bibinfo{title}{Resonant raman spectroscopy of twisted multilayer
  graphene}.
\newblock \emph{\bibinfo{journal}{Nature Communications}}
  \textbf{\bibinfo{volume}{5}}, \bibinfo{pages}{5309} (\bibinfo{year}{2014}).

\bibitem{del-nanolett-2016}
\bibinfo{author}{del Corro, E.} \emph{et~al.}
\newblock \bibinfo{title}{{Atypical Exciton¨CPhonon Interactions in WS$_2$ and
  WSe$_2$ Monolayers Revealed by Resonance Raman Spectroscopy}}.
\newblock \emph{\bibinfo{journal}{Nano Letters}} \textbf{\bibinfo{volume}{16}},
  \bibinfo{pages}{2363--2368} (\bibinfo{year}{2016}).

\bibitem{CBR-Naturecom-2017}
\bibinfo{author}{Carvalho, B.~R.} \emph{et~al.}
\newblock \bibinfo{title}{{Intervalley scattering by acoustic phonons in
  two-dimensional MoS$_2$ revealed by double-resonance Raman spectroscopy}}.
\newblock \emph{\bibinfo{journal}{Nature Communications}}
  \textbf{\bibinfo{volume}{8}}, \bibinfo{pages}{14670} (\bibinfo{year}{2017}).

\bibitem{PhysRevLett-117-2016}
\bibinfo{author}{Singh, A.} \emph{et~al.}
\newblock \bibinfo{title}{{Long-Lived Valley Polarization of Intravalley Trions
  in Monolayer ${\mathrm{WSe}}_{2}$}}.
\newblock \emph{\bibinfo{journal}{Phys. Rev. Lett.}}
  \textbf{\bibinfo{volume}{117}}, \bibinfo{pages}{257402}
  (\bibinfo{year}{2016}).

\bibitem{WSe2-prl-2015}
\bibinfo{author}{Wang, G.} \emph{et~al.}
\newblock \bibinfo{title}{{Double Resonant Raman Scattering and Valley
  Coherence Generation in Monolayer ${\mathrm{WSe}}_{2}$}}.
\newblock \emph{\bibinfo{journal}{Phys. Rev. Lett.}}
  \textbf{\bibinfo{volume}{115}}, \bibinfo{pages}{117401}
  (\bibinfo{year}{2015}).

\bibitem{cu2o-prb-1973}
\bibinfo{author}{Williams, P.~F.} \& \bibinfo{author}{Porto, S. P.~S.}
\newblock \bibinfo{title}{{Symmetry-Forbidden Resonant Raman Scattering in
  ${\mathrm{Cu}}_{2}$O}}.
\newblock \emph{\bibinfo{journal}{Phys. Rev. B}} \textbf{\bibinfo{volume}{8}},
  \bibinfo{pages}{1782--1785} (\bibinfo{year}{1973}).

\bibitem{cu2o-prl-1973}
\bibinfo{author}{Yu, P.~Y.}, \bibinfo{author}{Shen, Y.~R.},
  \bibinfo{author}{Petroff, Y.} \& \bibinfo{author}{Falicov, L.~M.}
\newblock \bibinfo{title}{{Resonance Raman Scattering at the Forbidden Yellow
  Exciton in ${\mathrm{Cu}}_{2}$O}}.
\newblock \emph{\bibinfo{journal}{Phys. Rev. Lett.}}
  \textbf{\bibinfo{volume}{30}}, \bibinfo{pages}{283--286}
  (\bibinfo{year}{1973}).

\bibitem{Yeyu-Natphoton-2015}
\bibinfo{author}{Ye, Y.} \emph{et~al.}
\newblock \bibinfo{title}{Monolayer excitonic laser}.
\newblock \emph{\bibinfo{journal}{Nat Photon}} \textbf{\bibinfo{volume}{9}},
  \bibinfo{pages}{733--737} (\bibinfo{year}{2015}).

\bibitem{trion-NC-2016}
\bibinfo{author}{Plechinger, G.} \emph{et~al.}
\newblock \bibinfo{title}{Trion fine structure and coupled spin-valley dynamics
  in monolayer tungsten disulfide}.
\newblock \emph{\bibinfo{journal}{Nature Communications}}
  \textbf{\bibinfo{volume}{7}}, \bibinfo{pages}{12715} (\bibinfo{year}{2016}).

\bibitem{ZhangXX-prl-2015}
\bibinfo{author}{Zhang, X.-X.}, \bibinfo{author}{You, Y.},
  \bibinfo{author}{Zhao, S. Y.~F.} \& \bibinfo{author}{Heinz, T.~F.}
\newblock \bibinfo{title}{{Experimental Evidence for Dark Excitons in Monolayer
  ${\mathrm{WSe}}_{2}$}}.
\newblock \emph{\bibinfo{journal}{Phys. Rev. Lett.}}
  \textbf{\bibinfo{volume}{115}}, \bibinfo{pages}{257403}
  (\bibinfo{year}{2015}).

\bibitem{Fano-pr-1961}
\bibinfo{author}{Fano, U.}
\newblock \bibinfo{title}{Effects of configuration interaction on intensities
  and phase shifts}.
\newblock \emph{\bibinfo{journal}{Phys. Rev.}} \textbf{\bibinfo{volume}{124}},
  \bibinfo{pages}{1866--1878} (\bibinfo{year}{1961}).

\bibitem{RevModPhys-fano-review}
\bibinfo{author}{Miroshnichenko, A.~E.}, \bibinfo{author}{Flach, S.} \&
  \bibinfo{author}{Kivshar, Y.~S.}
\newblock \bibinfo{title}{Fano resonances in nanoscale structures}.
\newblock \emph{\bibinfo{journal}{Rev. Mod. Phys.}}
  \textbf{\bibinfo{volume}{82}}, \bibinfo{pages}{2257--2298}
  (\bibinfo{year}{2010}).

\bibitem{FANORESONANCE-NM-2010}
\bibinfo{author}{Luk'yanchuk, B.} \emph{et~al.}
\newblock \bibinfo{title}{The {Fano} resonance in plasmonic nanostructures and
  metamaterials}.
\newblock \emph{\bibinfo{journal}{Nat Mater}} \textbf{\bibinfo{volume}{9}},
  \bibinfo{pages}{707--715} (\bibinfo{year}{2010}).

\bibitem{semifano-prb-1975}
\bibinfo{author}{Balkanski, M.}, \bibinfo{author}{Jain, K.~P.},
  \bibinfo{author}{Beserman, R.} \& \bibinfo{author}{Jouanne, M.}
\newblock \bibinfo{title}{{Theory of interference distortion of Raman
  scattering line shapes in semiconductors}}.
\newblock \emph{\bibinfo{journal}{Phys. Rev. B}} \textbf{\bibinfo{volume}{12}},
  \bibinfo{pages}{4328--4337} (\bibinfo{year}{1975}).

\bibitem{nanotube-nature-1997}
\bibinfo{author}{Rao, A.~M.}, \bibinfo{author}{Eklund, P.~C.},
  \bibinfo{author}{Bandow, S.} \& \bibinfo{author}{Thess, R.~E.,
  A.and~Smalley}.
\newblock \bibinfo{title}{Evidence for charge transfer in doped carbon nanotube
  bundles from raman scattering}.
\newblock \emph{\bibinfo{journal}{Nature}} \textbf{\bibinfo{volume}{388}},
  \bibinfo{pages}{257--259} (\bibinfo{year}{1997}).

\bibitem{PhysRevLett-graphene-2009}
\bibinfo{author}{Kuzmenko, A.~B.} \emph{et~al.}
\newblock \bibinfo{title}{Gate tunable infrared phonon anomalies in bilayer
  graphene}.
\newblock \emph{\bibinfo{journal}{Phys. Rev. Lett.}}
  \textbf{\bibinfo{volume}{103}}, \bibinfo{pages}{116804}
  (\bibinfo{year}{2009}).

\bibitem{ZJ-nanolett-2011}
\bibinfo{author}{Zhang, J.} \emph{et~al.}
\newblock \bibinfo{title}{{Raman Spectroscopy of Few-Quintuple Layer
  Topological Insulator Bi$_2$Se$_3$ Nanoplatelets}}.
\newblock \emph{\bibinfo{journal}{Nano Letters}} \textbf{\bibinfo{volume}{11}},
  \bibinfo{pages}{2407--2414} (\bibinfo{year}{2011}).

\bibitem{RevModPhys-79-75}
\bibinfo{author}{Devereaux, T.~P.} \& \bibinfo{author}{Hackl, R.}
\newblock \bibinfo{title}{Inelastic light scattering from correlated
  electrons}.
\newblock \emph{\bibinfo{journal}{Rev. Mod. Phys.}}
  \textbf{\bibinfo{volume}{79}}, \bibinfo{pages}{175--233}
  (\bibinfo{year}{2007}).

\bibitem{Mark-F-prl-2010}
\bibinfo{author}{Mak, K.~F.}, \bibinfo{author}{Lee, C.}, \bibinfo{author}{Hone,
  J.}, \bibinfo{author}{Shan, J.} \& \bibinfo{author}{Heinz, T.~F.}
\newblock \bibinfo{title}{Atomically thin {MoS$_2$}: A new direct-gap
  semiconductor}.
\newblock \emph{\bibinfo{journal}{Phys. Rev. Lett.}}
  \textbf{\bibinfo{volume}{105}}, \bibinfo{pages}{136805}
  (\bibinfo{year}{2010}).

\bibitem{LYL-PRB-2014}
\bibinfo{author}{Li, Y.} \emph{et~al.}
\newblock \bibinfo{title}{Measurement of the optical dielectric function of
  monolayer transition-metal dichalcogenides:{ ${\mathrm{MoS}}_{2}$,
  $\mathrm{Mo}\mathrm{S}{\mathrm{e}}_{2}$, ${\mathrm{WS}}_{2}$, and
  $\mathrm{WS}{\mathrm{e}}_{2}$}}.
\newblock \emph{\bibinfo{journal}{Phys. Rev. B}} \textbf{\bibinfo{volume}{90}},
  \bibinfo{pages}{205422} (\bibinfo{year}{2014}).

\bibitem{Zhang-prb-2013}
\bibinfo{author}{Zhang, X.} \emph{et~al.}
\newblock \bibinfo{title}{Raman spectroscopy of shear and layer breathing modes
  in multilayer {MoS$_2$}}.
\newblock \emph{\bibinfo{journal}{Phys. Rev. B.}}
  \textbf{\bibinfo{volume}{87}}, \bibinfo{pages}{115413}
  (\bibinfo{year}{2013}).

\bibitem{lixl-nt-2016}
\bibinfo{author}{Li, X.-L.} \emph{et~al.}
\newblock \bibinfo{title}{Determining layer number of two-dimensional flakes of
  transition-metal dichalcogenides by the raman intensity from substrates}.
\newblock \emph{\bibinfo{journal}{Nanotechnology}}
  \textbf{\bibinfo{volume}{27}}, \bibinfo{pages}{145704}
  (\bibinfo{year}{2016}).

\bibitem{ShiW-2D-2016}
\bibinfo{author}{Shi, W.} \emph{et~al.}
\newblock \bibinfo{title}{{Raman and photoluminescence spectra of
  two-dimensional nanocrystallites of monolayer $WS_2$ and $WSe_2$}}.
\newblock \emph{\bibinfo{journal}{2D Materials}} \textbf{\bibinfo{volume}{3}},
  \bibinfo{pages}{025016} (\bibinfo{year}{2016}).

\bibitem{ZhangYB-NN-2010}
\bibinfo{author}{Tang, T.-T.} \emph{et~al.}
\newblock \bibinfo{title}{A tunable phonon-exciton {Fano} system in bilayer
  graphene}.
\newblock \emph{\bibinfo{journal}{Nat Nano}} \textbf{\bibinfo{volume}{5}},
  \bibinfo{pages}{32--36} (\bibinfo{year}{2010}).

\bibitem{Abstreiter1984}
\bibinfo{author}{Abstreiter, G.}, \bibinfo{author}{Cardona, M.} \&
  \bibinfo{author}{Pinczuk, A.}
\newblock \emph{\bibinfo{title}{{Light Scattering in Solids IV: Electronics
  Scattering, Spin Effects, SERS, and Morphic Effects}}}
  (\bibinfo{publisher}{Springer Berlin Heidelberg}, \bibinfo{address}{Berlin,
  Heidelberg}, \bibinfo{year}{1984}).

\bibitem{Phonondispersion-2011}
\bibinfo{author}{Molina-S\'anchez, A.} \& \bibinfo{author}{Wirtz, L.}
\newblock \bibinfo{title}{{Phonons in single-layer and few-layer MoS${}_{2}$
  and WS${}_{2}$}}.
\newblock \emph{\bibinfo{journal}{Phys. Rev. B}} \textbf{\bibinfo{volume}{84}},
  \bibinfo{pages}{155413} (\bibinfo{year}{2011}).

\bibitem{Ferrari-200747}
\bibinfo{author}{Ferrari, A.~C.}
\newblock \bibinfo{title}{{Raman spectroscopy of graphene and graphite:
  Disorder, electron¨Cphonon coupling, doping and nonadiabatic effects}}.
\newblock \emph{\bibinfo{journal}{Solid State Communications}}
  \textbf{\bibinfo{volume}{143}}, \bibinfo{pages}{47 -- 57}
  (\bibinfo{year}{2007}).

\bibitem{linewidth-prb-1984}
\bibinfo{author}{Men\'endez, J.} \& \bibinfo{author}{Cardona, M.}
\newblock \bibinfo{title}{{Temperature dependence of the first-order Raman
  scattering by phonons in Si, Ge, and
  $\ensuremath{\alpha}-\mathrm{S}\mathrm{n}$: Anharmonic effects}}.
\newblock \emph{\bibinfo{journal}{Phys. Rev. B}} \textbf{\bibinfo{volume}{29}},
  \bibinfo{pages}{2051--2059} (\bibinfo{year}{1984}).

\bibitem{RICHTER-SSC-1981}
\bibinfo{author}{Richter, H.}, \bibinfo{author}{Wang, Z.} \&
  \bibinfo{author}{Ley, L.}
\newblock \bibinfo{title}{The one phonon raman spectrum in microcrystalline
  silicon}.
\newblock \emph{\bibinfo{journal}{Solid State Communications}}
  \textbf{\bibinfo{volume}{39}}, \bibinfo{pages}{625 -- 629}
  (\bibinfo{year}{1981}).

\bibitem{Froehlicher-nanoletter-2015}
\bibinfo{author}{Froehlicher, G.} \emph{et~al.}
\newblock \bibinfo{title}{{Unified Description of the Optical Phonon Modes in
  N-Layer MoTe$_2$}}.
\newblock \emph{\bibinfo{journal}{Nano Letters}} \textbf{\bibinfo{volume}{15}},
  \bibinfo{pages}{6481--6489} (\bibinfo{year}{2015}).

\bibitem{Tonndorf-oe-2013}
\bibinfo{author}{Tonndorf, P.} \emph{et~al.}
\newblock \bibinfo{title}{{Photoluminescence emission and Raman response of
  monolayer MoS$_2$, MoSe$_2$, and WSe$_2$}}.
\newblock \emph{\bibinfo{journal}{Opt. Express}} \textbf{\bibinfo{volume}{21}},
  \bibinfo{pages}{4908--4916} (\bibinfo{year}{2013}).

\bibitem{kim-nanoletter-2016}
\bibinfo{author}{Kim, K.}, \bibinfo{author}{Lee, J.-U.}, \bibinfo{author}{Nam,
  D.} \& \bibinfo{author}{Cheong, H.}
\newblock \bibinfo{title}{{Davydov Splitting and Excitonic Resonance Effects in
  Raman Spectra of Few-Layer MoSe$_2$}}.
\newblock \emph{\bibinfo{journal}{ACS Nano}} \textbf{\bibinfo{volume}{10}},
  \bibinfo{pages}{8113--8120} (\bibinfo{year}{2016}).

\bibitem{Staiger-prb-2015}
\bibinfo{author}{Staiger, M.} \emph{et~al.}
\newblock \bibinfo{title}{{Splitting of monolayer out-of-plane ${A}_{1}^{'}$
  Raman mode in few-layer ${WS}_{2}$}}.
\newblock \emph{\bibinfo{journal}{Phys. Rev. B}} \textbf{\bibinfo{volume}{91}},
  \bibinfo{pages}{195419} (\bibinfo{year}{2015}).

\bibitem{PhysRevLett-26-86}
\bibinfo{author}{Martin, R.~M.} \& \bibinfo{author}{Damen, T.~C.}
\newblock \bibinfo{title}{{Breakdown of Selection Rules in Resonance Raman
  Scattering}}.
\newblock \emph{\bibinfo{journal}{Phys. Rev. Lett.}}
  \textbf{\bibinfo{volume}{26}}, \bibinfo{pages}{86--88}
  (\bibinfo{year}{1971}).

\bibitem{cu2o-prl-1974}
\bibinfo{author}{Yu, P.~Y.} \& \bibinfo{author}{Shen, Y.~R.}
\newblock \bibinfo{title}{{Multiple Resonance Effects on Raman Scattering at
  the Yellow-Exciton Series of ${\mathrm{Cu}}_{2}$O}}.
\newblock \emph{\bibinfo{journal}{Phys. Rev. Lett.}}
  \textbf{\bibinfo{volume}{32}}, \bibinfo{pages}{373--376}
  (\bibinfo{year}{1974}).

\bibitem{PhysRevLett-3141-1973}
\bibinfo{author}{Compaan, A.} \& \bibinfo{author}{Cummins, H.~Z.}
\newblock \bibinfo{title}{{Resonant Quadrupole-Dipole Raman Scattering at the
  $1S$ Yellow Exciton in ${\mathrm{Cu}}_{2}$O}}.
\newblock \emph{\bibinfo{journal}{Phys. Rev. Lett.}}
  \textbf{\bibinfo{volume}{31}}, \bibinfo{pages}{41--44}
  (\bibinfo{year}{1973}).

\bibitem{Splitting-PRB-2016}
\bibinfo{author}{Echeverry, J.~P.}, \bibinfo{author}{Urbaszek, B.},
  \bibinfo{author}{Amand, T.}, \bibinfo{author}{Marie, X.} \&
  \bibinfo{author}{Gerber, I.~C.}
\newblock \bibinfo{title}{Splitting between bright and dark excitons in
  transition metal dichalcogenide monolayers}.
\newblock \emph{\bibinfo{journal}{Phys. Rev. B}} \textbf{\bibinfo{volume}{93}},
  \bibinfo{pages}{121107} (\bibinfo{year}{2016}).

\bibitem{DMS-1984}
\bibinfo{author}{Gene, D. M. S.~D.}, \bibinfo{author}{Jorio, A.} \&
  \bibinfo{author}{Heine, V.}
\newblock \emph{\bibinfo{title}{{Group Theory: Application to the Physics of
  Condensed Matter}}}, vol.~\bibinfo{volume}{61}.

\bibitem{Xiong-arXiv-2017}
\bibinfo{author}{Jiang, C.} \emph{et~al.}
\newblock \bibinfo{title}{Microsecond dark-exciton valley polarization memory
  in {2$D$ heterostructures}}.
\newblock \emph{\bibinfo{journal}{arXiv:1703.03133}}  (\bibinfo{year}{2017}).

\bibitem{XD-prl-2012}
\bibinfo{author}{Xiao, D.}, \bibinfo{author}{Liu, G.-B.},
  \bibinfo{author}{Feng, W.}, \bibinfo{author}{Xu, X.} \& \bibinfo{author}{Yao,
  W.}
\newblock \bibinfo{title}{{Coupled Spin and Valley Physics in Monolayers of
  ${\mathrm{MoS}}_{2}$ and Other Group-VI Dichalcogenides}}.
\newblock \emph{\bibinfo{journal}{Phys. Rev. Lett.}}
  \textbf{\bibinfo{volume}{108}}, \bibinfo{pages}{196802}
  (\bibinfo{year}{2012}).

\bibitem{NC-darkexciton-2017}
\bibinfo{author}{Feierabend, M.}, \bibinfo{author}{Bergh$\ddot{a}$user, G.},
  \bibinfo{author}{Knorr, A.} \& \bibinfo{author}{Malic, E.}
\newblock \bibinfo{title}{Proposal for dark exciton based chemical sensors}.
\newblock \emph{\bibinfo{journal}{Nature Communications}}
  \textbf{\bibinfo{volume}{8}}, \bibinfo{pages}{14776} (\bibinfo{year}{2017}).

\end{thebibliography}
\end{document}